\documentstyle[amssymb,epsfig]{new}
\newcommand{\be}{\begin{equation}}
\newcommand{\ee}{\end{equation}}
\def\vev#1{\left\langle #1\right\rangle}
\begin{document}
%
%IC/98 
% 
% 
%
\title{Neutrino Conversion and Neutrino\\ 
Astrophysics\footnote{Talk 
given  at the Symposium
``New Era in Neutrino Physics",
Tokyo Metropolitan University,
Japan, 11-12 June, 1998.}}
\author{A. Yu. Smirnov\\
{\it International Center for Theoretical Physics, Strada Costiera 11, 
34100 Trieste, Italy and\\   
Institute for Nuclear Research, RAN, Moscow, Russia}}

\maketitle

\section*{Abstract}

We consider main ingredients which determine neutrino transformations in
media. Strong transformations relevant for the astrophysics  
can be due to large depth  oscillations, resonance 
conversion, parametric resonance effect, interplay of oscillations 
and inelastic collisions. 
Properties of transitions are discussed using
the graphic representation. 
The applications of the transitions to supernova neutrinos
are described. 
The supernova neutrinos can probe whole neutrino mass spectrum. 
Their studies will help to 
identify the pattern of neutrino  mass and mixing.

\section{Introduction}
\label{Introduction}

The effects of neutrino propagation are determined by 
the following three ingredients: \\ 
(i) Properties of media (physical conditions):  density,  chemical
composition, 
polarization, motion;\\  
(ii) Density profiles: effective density distribution on the  
way of neutrinos;\\ 
(iii) Pattern of neutrino masses and mixing.

Variety of the physical conditions, profiles and possible mass spectra 
of neutrinos determines a large number of effects which can be observed by
present and future experiments.

In this review we will consider the system of three (or more) 
neutrinos $\nu_f = (\nu_e, \nu_{\mu}, \nu_{\tau} ...)$ 
mixed by mass terms (vacuum mixing). 
In the ultrarelativistic limit a propagation of these 
neutrinos is described by the evolution equation  \cite{W,MS} 
\be
i\displaystyle{\frac{d\nu_f}{dt}= H\nu_f}, ~~~
H\simeq\displaystyle{\frac{M^2}{2E} + V_f} ~, 
\label{evolution}
\ee
where 
\be
M = S M^{diag} S^{\dagger}
\label{mixi}
\ee  
is the mass matrix in flavor basis. 
Here $M^{diag} \equiv  diag(m_1, m_2, m_3, ...)$,  
$m_i$ are the masses of neutrinos. $S$ is the mixing matrix 
determined by $\nu_f = S \nu$, where  
$\nu = (\nu_1, \nu_2, \nu_3, ...) $ are 
the eigenstates of the mass matrix.  
$E$ is the neutrino energy. 
$V_f$ is the matrix of the effective potentials which describes the
interactions of neutrinos with  medium. The real part of the  
potential corresponds to refraction effect. The imaginary part 
describes inelastic collisions which lead to 
depart of the neutrino from the coherent  state. 
The ratio of the imaginary to  real parts of the potentials 
is $V_I/V_R \sim \sqrt{s}/ m_W \ll 1$ for low energy neutrinos,  
and  for  many applications one can neglect $V_I$.  
The Hamiltonian is hermitian in this case.   

For two neutrino  case, {\it e.g.} $(\nu_e, \nu_{\mu})$,  we get
explicitly: 
\be
H = \displaystyle{\frac{\Delta m^2}{4E}}\left(
\begin{array}{ccc}
- \cos 2\theta + 4 V_{e \mu} E/\Delta m^2 & \sin2 \theta \\
\sin 2 \theta & \cos 2\theta
\end{array}
\right)~, ~
\label{ham}
\ee
where $\theta$ is the mixing angle in vacuum, 
$\Delta m^2 \equiv m^2_2 - m^2_1$, 
and $V_{e \mu} \equiv V_e - V_{\mu}$.  

%%%%%%%%%%%%%%%%%%%%%%%%%%%%%%%%%%%%%%%%%%%%%%%%%%%%%%%%%%%%%%%%%%%%%%%%%
\section{Physical Conditions}
%%%%%%%%%%%%%%%%%%%%%%%%%%%%%%%%%%%%%%%%%%%%%%%%%%%%%%%%%%%%%%%%%%%%%%%%%

Properties of medium are described by the effective 
potential $V_f$ which can be calculated as 
\be
V_f = \langle \Psi| H_{int}| \Psi\rangle ~.
\label{matel}
\ee
Here $\Psi$ is the wave function of 
the system of neutrino and medium,
and $H_{int}$ is the Hamiltonian of interaction.
According to the standard model the matrix
of the potentials in the flavor basis,  $V_f$,
is practically diagonal: $V_f = diag(V_e, V_{\mu}, V_{\tau}, 0 ...)$.
Only difference of the diagonal elements is important.  
The  Hamiltonian $H_{int}$ is the effective four fermion Hamiltonian
due to exchange of the $W$ and $Z$ bosons: 
\be
H_{int} = \frac{G_F}{\sqrt{2}}\bar{\nu} \gamma^{\mu}(1 - \gamma_5) \nu
\left\{\bar{e} \gamma_{\mu}(g_V + g_A \gamma_5) e + 
\bar{p} \gamma_{\mu}(g_V^p + g_A^p \gamma_5) p + 
\bar{n} \gamma_{\mu}(g_V^n + g_A^n \gamma_5) n \right\},  
\ee
where $G_F$ is the Fermi coupling constant; 
$g_V$ and  $g_A$ are the vector and the axial 
vector coupling constants.

Let us consider the effect of scattering on electrons.
We define the vector of polarization of electrons as 
\be
\vec{\lambda}_e \equiv \omega_e^{\dagger}\vec{\sigma}\omega_e,  
\ee 
where $\omega_e$ is  the two component spinor. 
Suppose  electrons have some  density distribution
over the momentum,  $\vec{p}_e$ and 
polarization $\vec{\lambda}_e$:   
$$  
\frac{f(\vec{\lambda}_e,\vec{p}_e)}{(2\pi)^3}.  
$$
Then the  total number density  of electrons, $n_e$, equals
\be
\label{ne}
n_e = \sum_{\vec{\lambda}}
\int  \frac{d^3 p_e}{(2\pi)^3} ~f(\vec{\lambda_e},\vec{p}_e).
\ee
The average polarization of electrons is defined as
\be
\label{polave}
\langle \vec{\lambda}_{e}\rangle =
\frac{1}{n_e}\
\sum_{\vec{\lambda}}
\int  \frac{d^3 p_e}
{(2\pi)^3} \vec{\lambda} ~
f(\vec{\lambda}_e,\vec{p}_e). 
\label{avpol}
\ee
The matrix element (\ref{matel}) can be calculated as 
\be 
\sum_{\vec{\lambda}} \int  \frac{d^3 p_e}
{(2\pi)^3}  ~
f(\vec{\lambda}_e,\vec{p}_e) 
\langle e_{p, \lambda}|\bar{e} \gamma_{\mu}(g_V + g_A \gamma_5) e  
     | e_{p, \lambda}\rangle ,
\label{polave} 
\ee
where for free electrons $ |e_{p, \lambda}\rangle $ is the 
solution of the Dirac equation.\\ 

Let us consider the results of calculations of the 
potentials for the most important cases 
(for recent discussion see \cite{valle}).

%%%%%%%%%%%%%%%%%%%%%%%%%%%%%%%%%%%%%%%%%%%%%%%%%%%%%%%%%%%%%%%%%%%%%%
\subsection{Unpolarized medium} 
%%%%%%%%%%%%%%%%%%%%%%%%%%%%%%%%%%%%%%%%%%%%%%%%%%%%%%%%%%%%%%%%%%%%%%%

In this case $\vec{\lambda}_e = 0$,  the vector current only 
contributes  to the potential: 
\be
\label{potvec}
V = V^V(\vec{p}_e)  = \sqrt{2} G_F\, g_V  \frac{f_e (\vec{p}_e)}{(2\pi)^3}
\left(1  - \frac{\vec{p}_e \cdot \widehat{k}_{\nu}}{E_e} \right) ,
\label{unpol}
\ee
where $\widehat{k}_{\nu} \equiv \vec{p}_{\nu}/|\vec{p}_{\nu}|$ with
$\vec{p}_{\nu}$ being the neutrino momentum, $E_e$ is the energy of
electrons. 
For non-relativistic electrons (as well as for isotropic distribution 
of the ultrarelativistic electrons) only   
$\gamma^0$ component of the vector current gives non-zero effect 
and its matrix element equals  the density of electrons, $n_e$. Therefore 
the integration of (\ref{unpol}) over $\vec{p}$ leads to \cite{W}
\be
V  = \sqrt{2} G_F n_e g_V~.
\ee

In the case of moving medium also space components 
of the vector current give non-zero contribution: 
$
\langle \psi_e | \vec{\gamma}|\psi_e\rangle \propto \vec{v}
$
and \cite{lang,valle}
\be 
V  = \sqrt{2} G_F n_e g_V ( 1 - v\cdot \cos \beta)~,
\label{vpot}
\ee
where $\beta$ is the angle between the momenta of the electrons
and neutrino.
In the case of isotropic distribution the second term in  
(\ref{vpot}) disappears.

%%%%%%%%%%%%%%%%%%%%%%%%%%%%%%%%%%%%%%%%%%%%%%%%%%%%%%%%%%%%%%%%%%
\subsection{Polarized medium} 
%%%%%%%%%%%%%%%%%%%%%%%%%%%%%%%%%%%%%%%%%%%%%%%%%%%%%%%%%%%%%%%%%%

Now  the axial vector current, $\vec{\gamma} \gamma_5$,
also gives the contribution which is proportional
to the polarization of electrons~\cite{lang,valle}. In the 
{\it non-relativistic} case we get 
\be
\label{potaxnonr}
V^A  \approx - \sqrt{2} G_F\, g_A  n_e
(\hat{k}_{\nu} \cdot \langle \vec{\lambda}_e \rangle~) ,
\ee
where the average polarization of electrons is defined in (\ref{avpol}). 
In the case of {\it ultra-relativistic electrons} we find  
\be
\label{potaxur}
V^A  \approx \sqrt{2} G_F\, g_A
\frac{f(\vec{\lambda}_e,\vec{p}_e)}{(2\pi)^3}
(\widehat{k}_e \cdot \vec{\lambda}_e)
\left[1 - (\hat{k}_e \cdot \hat{k}_{\nu})
\right] ~.
\label{vax}
\ee
Here $\hat{k}_{e} \equiv \vec{p}_{e}/|\vec{p}_{e}|$. 
If electrons are polarized in the transverse plane the potential
is zero for any momenta of neutrinos.
The potential is suppressed if neutrinos and electrons
are moving in the same direction.
 
Suppose, the two equal electron fluxes move 
in opposite directions but have the same polarization  
along the momentum. Such a configuration 
is realized in the  {\it magnetized medium}
 (electrons in the lowest Landau level). 
For this case we find using (\ref{vax}): 
\be
\label{potaxlan}
V^A  = - \sqrt{2} G_F\, g_A  n_e
(\hat{k}_{\nu} \cdot \vec{\lambda}_e) .
\ee
Here $n_e$ is the total concentration of electrons in both
fluxes. Notice that this relativistic expression
coincides with the non-relativistic formula (\ref{potaxnonr}).

The total effective potential resulting from 
neutrino scattering on electrons,   
protons and neutrons in an electrically neutral medium can be
written in the form: 
\be
\label{pottotal}
V  = \sqrt{2} G_F\, n_e \left[ g_V - g_A \hat{k}_{\nu} \cdot
\langle{\vec{\lambda}_e\,}\rangle \right] + \sqrt{2} G_F\, n_n g_V^n~.
\label{vtot}
\ee
Here, the second term describes neutrino-nucleon 
scattering  with $n_n$ being the neutron concentration.   
Notice that in the electrically neutral medium 
the neutral current contributions from the neutrino-proton and the 
neutrino-electron  scattering cancel each other 
%(electrons and protons have opposite weak charges). 
and only neutrons contribute. Furthermore,  
in the case of unpolarized neutrons at rest, the  zeroth component of 
vector current contributes to the potential only.   

%%%%%%%%%%%%%%%%%%%%%%%%%%%%%%%%%%%%%%%%%%%%%%%%%%%%%%%%%%%%%
\subsection{Difference of potentials} 
%%%%%%%%%%%%%%%%%%%%%%%%%%%%%%%%%%%%%%%%%%%%%%%%%%%%%%%%%%%%%
 
Effect of a  medium on neutrino propagation is determined by the
difference of potentials. 
For  $\nu_e \rightarrow \nu_\mu, \nu_\tau$
(flavor) conversion only the charge current neutrino-electron scattering
gives a net contribution. 
Taking $g_V = - g_A = 1$ one finds from (\ref{vtot})
\be
\label{potflavor}
V_{e\mu}  = \sqrt{2} G_F\, n_e \left[ 1 +  \hat{k}_{\nu} \cdot
\langle {\vec{\lambda}_e\,}\rangle \right] 
=  \sqrt{2} G_F\, n_e \left[ 1 +  \vev{\lambda_e}\cos \alpha
\right]~, 
\ee
where $\alpha$ is the angle between the neutrino momentum and   
the averaged polarization vector of electrons. There is no effect of
nucleons in this case. Depending on the direction of
polarization the axial term can either enhance 
or suppress the potential. 
The maximal effect is obtained in the case of complete polarization
in the direction of  the neutrino momentum, $\vev{\lambda_e} = 1$.
In the case of complete polarization against the
neutrino momentum, $\cos \alpha = -1$, $\vev{\lambda_e} = 1$,
we get $V_{e\mu}=0$. 
Clearly, the axial vector term can not overcome the vector term,
$| V_V| \geq |V_A| $. 

In the case of conversion into sterile neutrinos, 
the difference of potentials gets also the contribution 
from the neutrino-nucleon scattering. If 
nucleons are  unpolarized, we find for $\nu_e - \nu_{s}$ system: 
\be 
\label{potes}
V_{es}  = \sqrt{2} G_F\, n_e \left[ \left(1- \frac{n_n}{2n_e} \right)
+ \frac{1}{2}\,\hat{k}_{\nu} \cdot \langle{\vec{\lambda}_e\,}\rangle 
\right].
\label{ester}
\ee
Now the polarization term can be bigger than the vector current one,
thus leading to the possibility of level crossing induced by the
axial term. The latter may have some implication to the 
neutrinos in the central parts of supernova. 

For the $\nu_{\mu} - \nu_{s}$ system we get: 
\be 
\label{potesm}
V_{\mu s}  = \sqrt{2} G_F\, \left(- \frac{n_n}{2} 
- \frac{n_e}{2}~ \hat{k}_{\nu} \cdot \langle{\vec{\lambda}_e}\rangle 
\right).
\ee
This result can be applied to the atmospheric neutrinos.   

%%%%%%%%%%%%%%%%%%%%%%%%%%%%%%%%%%%%%%%%%%%%%%%%%%%%%%%%%%%%%%%%%%%
%%%%%%%%%%%%%%%%%%%% ss2.4 %%%%%%%%%%%%%%%%%%%%%%%%%%%%%%%%%%%%%%%%
\subsection{Magnetized Medium}
%%%%%%%%%%%%%%%%%%%%%%%%%%%%%%%%%%%%%%%%%%%%%%%%%%%%%%%%%%%%%%%%%%%

In  presence of the  magnetic field the energy spectrum of electrons 
is quantized. 
It  consists of the lowest Landau level, $n=0, \lambda_z = -1$,
plus pairs of degenerate levels with opposite polarizations.
As a result, the contributions from all the levels 
but  the lowest one cancel each other  
\cite{disp,Raffelt,SemikozValle}. So, the average
polarization of electrons is determined by  $ n_0$ --  the electron number
density in the lowest Landau level:  
\be \label{spin2}
\vev{\lambda_{e}} =  - \frac{n_0}{n_e}~.  
\ee
In strongly degenerate electron gas
\be
\label{nzero}
n_0 \approx \frac{eB p_F}{2\pi^2}~,
\ee 
where
$p_F = \sqrt{\mu^2  - m^2_e}$ is the  Fermi momentum 
determined from the expression for
the total electron concentration $n_e$. 
In the weak field limit, 
$
eB \ll p_F^2, 
$
we get the usual expression for $p_F$  in a medium
without magnetic field: 
$ 
p_F \simeq (3\pi^2 n_e)^{1/3}~.
$
Inserting this $p_F$ in (\ref{nzero}) we have 
\be
\label{nz}
n_0 =  \frac{eB}{2}\left(\frac{3n_e}{\pi^4}\right)^{1/3} ~.
\ee 
Therefore   $\vev{\lambda_{e}}$     increases linearly with $B$
and decreases as $n_e^{-2/3}$. Using (\ref{potflavor}) and (\ref{spin2})
we get for the effective potential of the electrons 
\cite{disp,DOlivo}  
\be
\label{potax4}
V = \sqrt{2}G_F\,n_e - 
\frac{G_F\, eB}{\sqrt{2}} \left(\frac{3n_e}{\pi^4}\right)^{1/3}
\cos\alpha_B~, 
\ee
where $\alpha_B$ is the angle between neutrino momentum and the 
magnetic field. 
The second term may be important in the central parts of supernovae 
\cite{Kusenko}.

\subsection{Non-local corrections. Thermal medium} 

The motion of scatterers manifests  also through the
correction to the propagator of the vector boson:
$G_F \rightarrow G_F (1 + q^2_W /m_W^2)$,
where $q^2_W$ is the four momentum of the intermediate  boson squared. 
Consequently, 
$$
V \rightarrow V_0 (1 + q^2_W /m_W^2). 
$$
In thermal bath $q^2_W \sim T^2$,  and
one gets \cite{notz}
\be
V_T  \sim  \sqrt{2} G_F n_e  A \frac{T^2}{m_W^2} ~,
\label{term}
\ee
where $A$ is the constant which depends on the composition of plasma.
The crucial feature is that 
the thermal correction (\ref{term}),
$V_T$, has  the same  signs for neutrinos and antineutrinos.
This comes about from the following facts: (i)  
For $\nu e$-scattering the $W-$ exchange occurs in the $t$ channel, 
whereas  for $\bar{\nu} e$-scattering -- in the $s$ channel. 
Therefore $q^2_W < 0$ for neutrinos  and    $q^2_W > 0$
for antineutrinos. (ii) The currents of 
neutrinos and antineutrinos have opposite signs.  So, in the 
transition from neutrino channel to antineutrino channel the  amplitude 
changes the sign twice.

In  thermal bath with non-zero lepton charge  
(the Early Universe)  the potential can be written as 
\be
V  = \sqrt{2} G_F n_{\gamma}
\left( \Delta L + A \frac{T^2}{m_W^2}\right)~,
\label{euniv}
\ee
where $n_{\gamma}$ is the photon density,
$\Delta L = (n_L - n_{\bar{L}})/n_{\gamma}$
is the leptonic asymmetry and
$n_L$,  $n_{\bar{L}}$ are the concentrations of the
leptons and antileptons.    
In CP symmetric medium only thermal  correction survives  
and $V = V_0 A {T^2}/{m_W^2}$.

In the Early Universe the matter effects can be important for
oscillations into
sterile neutrinos. Matter influences differently the neutrino and
antineutrino
channels, so that transitions
$\nu_{\tau} \rightarrow \nu_s$ , and 
$\bar\nu_{\tau} \rightarrow \bar\nu_s$
can create the $\nu_{\tau} - \bar{\nu}_{\tau}$ asymmetry 
in the Universe \cite{foot}. 
The leptonic asymmetry  influences
the primordial nucleosyntesis. 
It can also suppress further the production of  sterile
neutrinos, so that the concentration of these neutrinos
is much smaller than the equilibrium concentration
even in the case  of large mixing angle
and large mass squared difference.

%%%%%%%%%%%%%%%%%%%%%%%%%% sec3 %%%%%%%%%%%%%%%%%%%%%%%%%%%%%%%%%%%%%%%%%%
%%%%%%%%%%%%%%%%%%%%%%%%%%%%%%%%%%%%%%%%%%%%%%%%%%%%%%%%%%%%%%%%%%%%%%%%%%
\section{Density Profiles}
%%%%%%%%%%%%%%%%%%%%%%%%%%%%%%%%%%%%%%%%%%%%%%%%%%%%%%%%%%%%%%%%%%%%%%%%

When transitions of neutrinos are strong? 
This question is especially relevant for  astrophysics,  
where the effects  have observable consequences 
provided  that  transition probabilities are of the order one 
(apart from a few exceptional cases). 
Physical conditions are described by the effective potentials.  
The result of conversion 
depends on the density  profile, that is, on the change  of   
the effective potential  on the way of neutrinos.  

We start here with some elements of dynamics  and then 
consider different profiles which lead to strong transitions. 

%%%%%%%%%%%%%%%%%%%%%%%%%%%%%%%%%%%%%%%%%%%%%%%%%%%%%%%%%%%%%%%%%%%%%
%%%%%%%%%%%%%%%%%%%%%%%%%%%%% ss3.1 %%%%%%%%%%%%%%%%%%%%%%%%%%%%%%%%%%
\subsection{Elements of dynamics}
%%%%%%%%%%%%%%%%%%%%%%%%%%%%%%%%%%%%%%%%%%%%%%%%%%%%%%%%%%%%%%%%%%%%%

Let us consider the evolution equation (\ref{evolution}) for 
two neutrino species $(\nu_e, \nu_{\mu})$. 
The Hamiltonian is the function of the electron density,  
and consequently, the time: $H = H(n_e(t))$. 
For a given moment of time $t$ we can introduce the instantaneous 
eigenstates of $H$, $\nu_{im}(t)$,  and eigenvalues of $H$, $E_{im}(t)$,  
$(i = 1, 2)$: 
$$
H(t) \nu_{im}(t) = E_{im}(t)\nu_{im}(t)~. 
$$ 
The  eigenstates are related to the flavor states as 
\be
\nu_f= S(\theta_m)\nu_H .
\label{flavmass}
\ee
This equality can be considered as the definition of the 
mixing matrix in medium,  $S(\theta_m)$,  
and the mixing angle in medium $\theta_m$. 
Both the mixing angle and the eigenvalues are  functions 
of electron density: $\theta_m = \theta_m (n_e)$,  
$E_{im} = E_{im}(n_e)$. For $\theta_m$ we have explicitly 
\be
\tan 2\theta_m = \frac{\sin 2\theta}
{\cos 2\theta - 2 \sqrt{2} G_F n_e E/\Delta m^2}~.
\label{mam}
\ee
The mixing becomes maximal ($\theta_m  = \pi/4$) at 
the resonance density 
\be
n_e^{R} = \frac{\Delta m^2}{2E} \frac{\cos 2\theta}{\sqrt{2} G_F}~.
\label{eq:resonance}
\ee
From (\ref{flavmass}) we find the inverse relation: 
\be
\nu_{1m} = \cos\theta_m \nu_e - \sin \theta_m \nu_{\mu}, ~~ 
\nu_{2m} = \cos\theta_m \nu_{\mu} - \sin \theta_m \nu_{e}. 
\label{flav}
\ee
According to (\ref{flav})  the mixing angle $\theta_m$ determines the
$\nu_e-$, 
$\nu_{\mu}-$ 
(that is, flavor) content of the neutrino eigenstates. 
When   $n_e$ changes with distance, 
 $\theta_m$ also changes according to 
(\ref{mam}). Then from (\ref{flav}) we get  
the following conclusion.  In  the nonuniform medium the 
flavors of the eigenstates change: they  
uniquely follow a change of the electron density. 
When density changes from 
$n_e \gg n_e^R$ to $n_e \ll n_e^R$  (and if the vacuum mixing is small),  
the flavor of the $\nu_{1m}$ changes from almost $\nu_{\mu}$ to $\nu_e$
and the flavor of $\nu_{2m}$ -- from $\nu_e$ to $\nu_{\mu}$ (fig. 1). 
%%%%%%%%%%%%%%%%%%%%%%%%%%%%%%%%%%%%%%%%%%%%%%%%%%%%%%%%%%%%%%%%%%
%%%%%%%%%%%%%%%%%%%%%%%%%%%  res %%%%%%%%%%%%%%%%%%%%%%%%%%%%%%%%
\begin{figure}[htb]
\hbox to \hsize{\hfil\epsfxsize=10cm\epsfbox{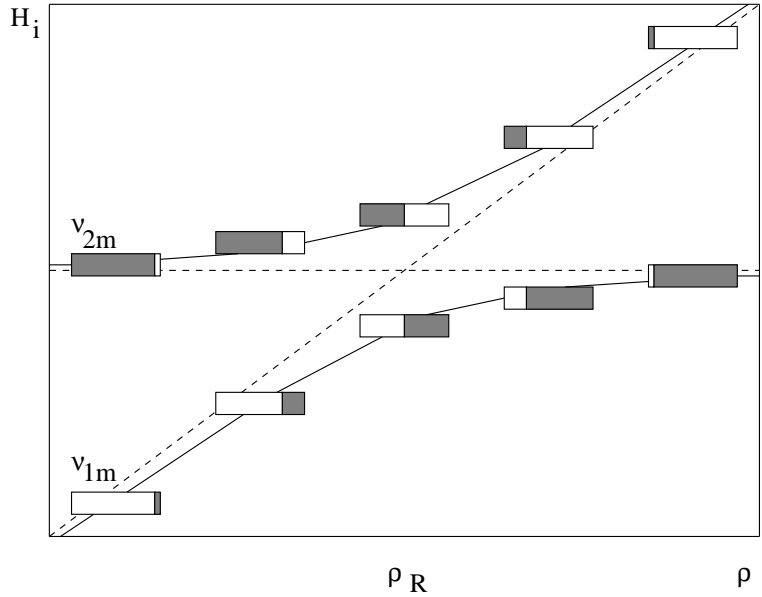}\hfil}
\caption{~~Energies  $H_i$ (solid lines) and flavors of the neutrino
eigenstates 
as functions of the effective density $\rho \equiv m_N n_e$  
($m_N$ is the mass of the nucleon).  
White  parts of boxes represent the electron flavor and 
shadowed  parts are the muon flavor.
} 
\end{figure}
%%%%%%%%%%%%%%%%%%%%%%%%%%%%%%%%%%%%%%%%%%%%%%%%%%%%%%%%%%%%%%%%%%
%%%%%%%%%%%%%%%%%%%%%%%%%%%%%%%%%%%%%%%%%%%%%%%%%%%%%%%%%%%%%%%%
%%%%%%%%%%%%%%%%%%%%% ss3.2 %%%%%%%%%%%%%%%%%%%%%%%%%%%%%%%%%%%%
\subsection{Degrees of freedom}
%%%%%%%%%%%%%%%%%%%%%%%%%%%%%%%%%%%%%%%%%%%%%%%%%%%%%%%%%%%%%%%%

An arbitrary neutrino state can be expressed in terms of the 
instantaneous eigenstates as 
\be
\nu (t) = \cos\theta_a \nu_{1m} + \sin \theta_a \nu_{2m} e^{i\phi}~,  
\label{exp}
\ee
where 
\begin{itemize}

\item
$\theta_a = \theta_a (t)$ determines  the admixtures of the
eigenstates in a given state;\\

\item
$\phi(t)$ is the phase difference between the two eigenstates (phase 
of oscillations): 
\be
\phi(t) = \int_0^t \Delta H dt' + \phi(t)_T~, 
\ee
where $\Delta H \equiv H_{1} - H_{2}$, 
the integral determines the adiabatic phase 
and $\phi(t)_T$ is the rest which can be related to 
violation of adiabaticity. It may also have a 
topological contribution  (Berry phase) in more complicated systems;\\ 

\item
Flavor content of the eigenstates  depends  
on time and 
changes according to  mixing angle change $\theta_m(n_e(t))$, 
as we have discussed in sect. 3.1. 

\end{itemize}

Different processes are associated with these three degrees of freedom.

%%%%%%%%%%%%%%%%%%%%%%%%%%%%%%%%%%%%%%%%%%%%%%%%%%%%%%%%%%%%%%%%%%%%
%%%%%%%%%%%%%%%%%%%%%%%%%% ss3.3 %%%%%%%%%%%%%%%%%%%%%%%%%%%%%%%%%%%
\subsection{Density matrix and Graphic representation}
%%%%%%%%%%%%%%%%%%%%%%%%%%%%%%%%%%%%%%%%%%%%%%%%%%%%%%%%%%%%%%%%%%%%

We will consider dynamics of transitions in different 
media using  graphic representation \cite{graph}. 
The representation is based on
analogy of the neutrino evolution with behaviour of
spin of the electron in the magnetic field. Indeed,
a neutrino state can be described by vector
\begin{equation}
\vec{\nu} = \left( {\rm Re} \nu_{e}^{\dagger} \nu_{\mu}, ~~
{\rm Im} \nu_{e}^{\dagger} \nu_{\mu},~~
\nu_{e}^{\dagger} \nu_{e} - 1/2 \right) ~,
\end{equation}
where $\nu_{i}$, ($i =  \mu, e$) are the neutrino wave functions.
(The elements of this vector are nothing but  components of the density
matrix.) 
Introducing  another  vector:
\begin{equation}
\vec{B} \equiv \frac{2 \pi}{l_m} (\sin 2 \theta_m,~~ 0 ,~~ \cos 2
\theta_m)~
\label{axisb}
\end{equation}
($l_m = 2\pi/\Delta H$ is the oscillation length in medium) which
corresponds to the magnetic field,  one gets from the evolution equations 
for the wave functions (\ref{evolution}) the  equation
\begin{equation}
\frac{d \vec{\nu}}{d t} = \left(\vec{B} \times \vec{\nu} \right)~.
\end{equation}  

The vector $\vec{\nu}$  moves (see fig.~2) 
on the surface of the cone with axis $\vec{B}$ 
according to increase of the oscillation phase, $\phi$.
The direction of the  axis, $\vec{B}$,
is determined by $2\theta_m$ (\ref{axisb}). 
%%%%%%%%%%%%%%%%%%%%%%%%%%%%%%%%%%%%%%%%%%%%%%%%%%%%%%%%%%%%%%%%%%%%%%%
%%%%%%%%%%%%%%%%%%%%%%%%%%% graf1a %%%%%%%%%%%%%%%%%%%%%%%%%%%%%%%%%%%
\begin{figure}[htb]
\hbox to \hsize{\hfil\epsfxsize=8cm\epsfbox{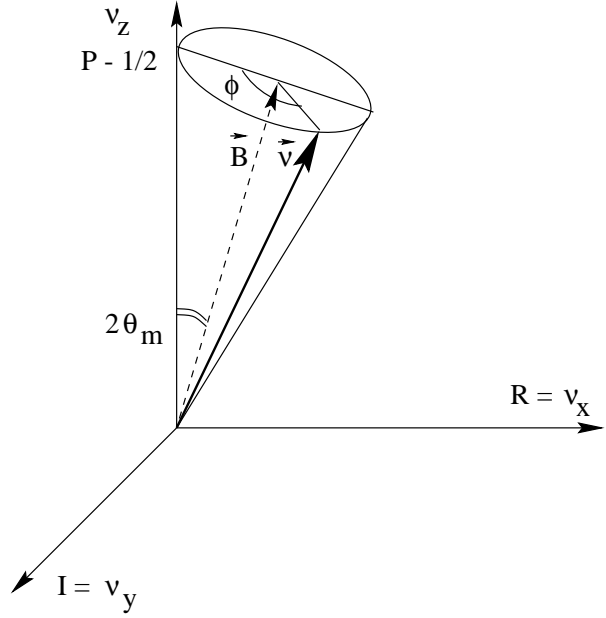}\hfil}
\caption{ Graphic representation of the neutrino 
oscillations in the uniform medium (see text). 
$R \equiv {\rm Re} \nu_{e}^{\dagger} \nu_{\mu}$, 
$I \equiv {\rm Im} \nu_{e}^{\dagger} \nu_{\mu}$,
$P \equiv \nu_{e}^{\dagger} \nu_{e}$. 
}
\end{figure}
%%%%%%%%%%%%%%%%%%%%%%%%%%%%%%%%%%%%%%%%%%%%%%%%%%%%%%%%%%%%%%%%%%%%%%%
The cone angle --    the angle between $\vec{\nu}$ and
$\vec{B}$ -- coincides with $2\theta_a$ (see (\ref{exp})).    
It depends 
both on mixing angle and on  the initial state, and in general,  changes
in
the process of evolution. 
If the  initial state is $\nu_{e}$,
the angle  equals  $2\theta_{a} =  2\theta_m$ in the initial moment. 
The projection  of $\vec{\nu}$ on the axis 
$\nu_{z}$, gives the probability
to find $\nu_{e}$ in a state $\vec{\nu}$:
\begin{equation}
P \equiv \nu_{e}^{\dagger} \nu_{e} =
\nu_z + \frac{1}{2} =
\cos^2 \frac{\theta_z}{2} ~.
\label{thetaz}
\end{equation}
Here $\nu_{z} \equiv 0.5 \cos \theta_z$,
and  $\theta_z$ is the angle between $\vec{\nu}$ and the axis $\nu_z$.

%%%%%%%%%%%%%%%%%%%%%%%%%%%%%%%%%%%%%%%%%%%%%%%%%%%%%%%%%%%%%%%%%%%%%%%
%%%%%%%%%%%%%%%%%%%%%%%%%%%%%% ss3.4 %%%%%%%%%%%%%%%%%%%%%%%%%%%%%%%%%% 
\subsection{Oscillations in the uniform medium}
%%%%%%%%%%%%%%%%%%%%%%%%%%%%%%%%%%%%%%%%%%%%%%%%%%%%%%%%%%%%%%%%%%%%%%%

In medium with constant density ($\theta_m = const$),
the evolution consists of $\vec{\nu}$- precession
around $\vec{B}$:  $\vec{\nu}$  moves monotonously 
according to increase of the oscillation phase, $\phi$. 
The evolution of neutrino  has a character of 
oscillations. Oscillations are the consequence of the monotonous 
change of the phase. Only this degree of freedom operates. 
Flavors of the eigenstates and the admixtures of the eigenstates in a
given state are fixed: 
\begin{equation}
\theta_m = const, ~~~~ \theta_a = const, ~~~~ \phi = (H_2  - H_1)t.  
\end{equation}

The mixing angle depends on the neutrino energy. Therefore 
for different energies vectors of the neutrino states 
will rotate around 
different axes with different cone angles. At the resonance energy, 
the rotation proceeds around $\nu_x$, and the cone angle equals $\pi/2$.  

Obviously, to get large transition effect one needs to have 
both large mixing angle and the phase of oscillations 
about  $\pi$: $\theta_m \sim \pi/4$, $\phi \sim \pi$. 
This condition  can be realized inside the Earth,  where the  
density profile can be
approximated by several layers with constant densities. 

%%%%%%%%%%%%%%%%%%%%%%%%%%%%%%%%%%%%%%%%%%%%%%%%%%%%%%%%%%%%%%%%%%%%%
%%%%%%%%%%%%%%%%%%%%%%% ss3.5 %%%%%%%%%%%%%%%%%%%%%%%%%%%%%%%%%%%%%%%
\subsection{Adiabatic resonance conversion} 
%%%%%%%%%%%%%%%%%%%%%%%%%%%%%%%%%%%%%%%%%%%%%%%%%%%%%%%%%%%%%%%%%%%%%

Suppose the density varies    
on the way of neutrinos 
({\it e.g.} decreases)  slowly. 
In this case the evolution is characterized by the  
following features: 

1. Mixing angle  changes according to  density change. 
Correspondingly, flavors of the eigenstates change. 

2. If $n_e$ changes slowly enough, so that
\be
|\dot{\theta}_m| \ll |H_2  - H_1|~,   
\label{adiab}
\ee
then in the first approximation  
the transitions $\nu_{1m} \leftrightarrow \nu_{2m}$ 
can be  neglected. The condition (\ref{adiab})
is called the {\it adiabaticity condition}.
In the adiabatic approximation (as in the cases of 
vacuum and uniform medium)  the eigenstates propagate independently.
This means that the  angle  $\theta_a$ is constant;  
admixtures of the eigenstates are
conserved.  

3. The phase between the eigenstates changes,  
leading to oscillations.  

Thus, the evolution in the adiabatic approximation is characterized by 
\begin{equation}
\theta_m = \theta_m (n_e), ~~~~ \theta_a = const,  ~~~
\phi = \int (H_2  - H_1) dt. 
\end{equation}

The interplay takes place between  the oscillations and the effects
related to the change of  flavors of the eigenstates. 
%%%%%%%%%%%%%%%%%%%%%%%%%%%%%%%%%%%%%%%%%%%%%%%%%%%%%%%%%%%%%%%%%
%%%%%%%%%%%%%%%%%%%%%%%%% graf2a.eps %%%%%%%%%%%%%%%%%%%%%%%%%%%%%%
\begin{figure}[htb]
\hbox to \hsize{\hfil\epsfxsize=6.5cm\epsfbox{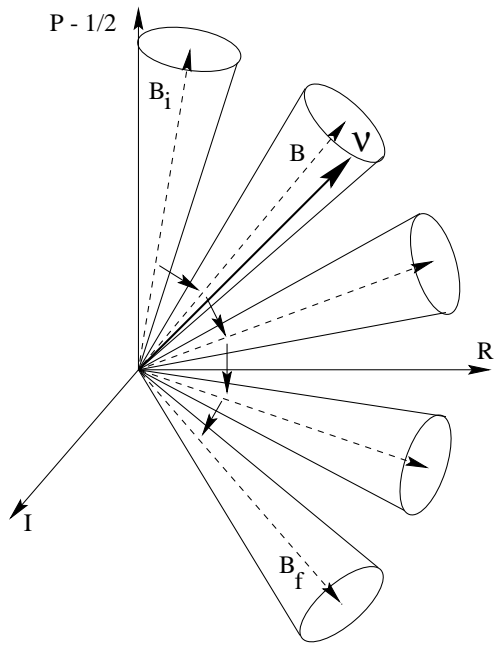}\hfil}
\caption{ Graphic representation of the neutrino  
adiabatic conversion (see text). 
}
\end{figure}
%%%%%%%%%%%%%%%%%%%%%%%%%%%%%%%%%%%%%%%%%%%%%%%%%%%%%%%%%%%%%%%%%%%%
%%%%%%%%%%%%%%%%%%%%%%%%%%%%%%%%%%%%%%%%%%%%%%%%%%%%%%%%%%%%%%%%%%%
Graphically, the adiabatic conversion  
is described in the following way (fig. 3).  
The axis of the cone rotates according to density change. 
The cone angle is unchanged (adiabaticity). The 
evolution consists of the  rotation of the cone and the rotation  
of the neutrino vector on the surface of the cone.

A strong transition occurs when the interval of density changes 
in sufficiently large. The initial density 
should be larger than 
the resonance density, whereas the final density 
should be smaller than 
$n_R$. 

%%%%%%%%%%%%%%%%%%%%%%%%%%%%%%%%%%%%%%%%%%%%%%%%%%%%%%%%%%%%%%%%%%%%%%%
%%%%%%%%%%%%%%%%%%%%%%%%%%%% ss3.6 %%%%%%%%%%%%%%%%%%%%%%%%%%%%%%%%%%%%
\subsection{Non-adiabatic resonance conversion}
%%%%%%%%%%%%%%%%%%%%%%%%%%%%%%%%%%%%%%%%%%%%%%%%%%%%%%%%%%%%%%%%%%%%%%%

Let us consider again a medium with monotonously changed density. 
If the density changes rapidly, the 
adiabaticity condition turns out to be broken
and the transitions  $\nu_{1m} \leftrightarrow \nu_{2m}$
become essential. This means that the admixtures of the eigenstates 
in a given state are changed, or equivalently, the angle 
$\theta_a$ is no more a constant. 
In this case all three degrees of freedom operate: 
\begin{equation}
\theta_m = \theta_m (n_e)~,  ~~~~ \theta_a = \theta_a(t), ~~~
\phi = \int (H_2  - H_1) dt.
\end{equation}

In the graphic  representation (fig.~4)   
the axis of the cone rotates according to density change.
The cone angle  changes (violation of the adiabaticity). 
The neutrino vector moves on the surface of the cone (phase change). 
%%%%%%%%%%%%%%%%%%%%%%%%%%%%%%%%%%%%%%%%%%%%%%%%%%%%%%%%%%%%%%%%%%%%
%%%%%%%%%%%%%%%%%%%%%%%%%%%%%%  graf3 %%%%%%%%%%%%%%%%%%%%%%%%%%%%%%
\begin{figure}[htb]
\hbox to \hsize{\hfil\epsfxsize=7cm\epsfbox{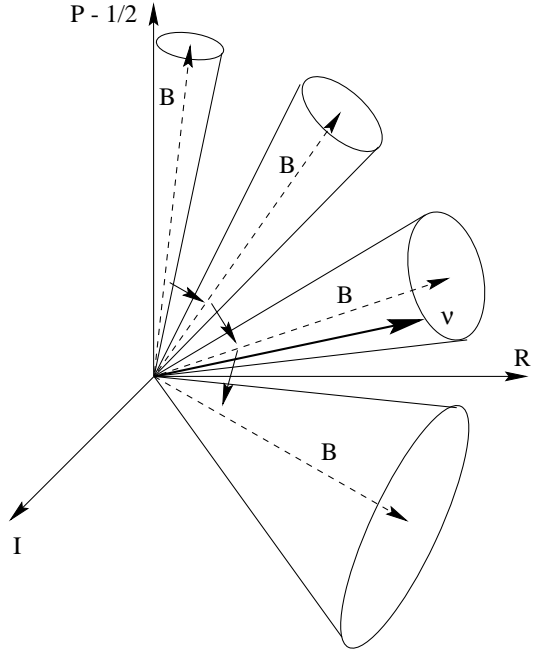}\hfil}
\caption{~~Graphic representation of the non-adiabatic neutrino  
conversion (see text). 
}
\end{figure}
%%%%%%%%%%%%%%%%%%%%%%%%%%%%%%%%%%%%%%%%%%%%%%%%%%%%%%%%%%%%%%%%%%%%
Typically, adiabaticity breaking leads to weakening of the  flavor  
transition.  Both adiabatic and non-adiabatic transitions can 
be  realized
inside the Sun and supernovae. 

%%%%%%%%%%%%%%%%%%%%%%%%%%%%%%%%%%%%%%%%%%%%%%%%%%%%%%%%%%%%%%%%%%%
%%%%%%%%%%%%%%%%%%%%%%%%%%  ss3.7 %%%%%%%%%%%%%%%%%%%%%%%%%%%%%%%%%
\subsection{Conversion  due to the parametric resonance}
%%%%%%%%%%%%%%%%%%%%%%%%%%%%%%%%%%%%%%%%%%%%%%%%%%%%%%%%%%%%%%%%%%%

Strong transitions discussed in sect 3.4 - 3.6  imply an existence of
the large 
effective mixing in whole medium (constant density) 
or at least in some layer (the resonance conversion).  
There is the way to get large transition without large 
(vacuum or even matter)  mixing. This can be done with periodically 
changed density \cite{ETC,Akh}. 

The simplest example is the so called ``castle wall" 
profile \cite{Akh}, when  the period $l_f$ is divided 
into two parts $l_1$ and $l_2$ ($l_1 + l_2 =  l_f$) 
and the density takes two different
values $n_1$ and $n_2$ in  parts $l_1$ and  $l_2$
respectively    
(in general $l_1 \neq l_2$). 

General condition of the parametric resonance is that 
the effective oscillation length equals the 
period of density perturbation, or \cite{plam} 
\begin{equation}
\int_{l_f} \frac{dx}{l_m} = k~,~~~~  (k = 1, 2, 3,...)~. 
\end{equation}
For the ``castle wall" profile the parametric resonance condition is
reduced to  equality of the oscillation phases acquired by 
neutrinos on the two parts of periods \cite{LS}:  
\begin{equation}
\Phi_1 = \Phi_2 = \pi ~.  
\label{pi}
\end{equation}
(The size of the layer equals half of the oscillation length in this
layer.)
%%%%%%%%%%%%%%%%%%%%%%%%%%%%%%%%%%%%%%%%%%%%%%%%%%%%%%%%%%%%%%%%%%%%%%%%
%%%%%%%%%%%%%%%%%%%%%%%%%%%%%%  graf4 %%%%%%%%%%%%%%%%%%%%%%%%%%%%%%%%%% 
\begin{figure}[htb]
\hbox to \hsize{\hfil\epsfxsize=7cm\epsfbox{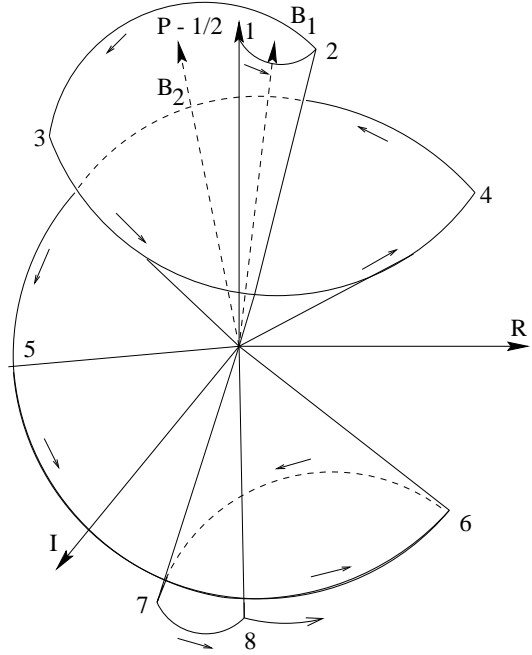}\hfil}
\caption{~~Graphic representation of the parametric 
enhancement effect. 
}
\end{figure}
%%%%%%%%%%%%%%%%%%%%%%%%%%%%%%%%%%%%%%%%%%%%%%%%%%%%%%%%%%%%%%%%%%%%%%%%
Graphic representation is shown in fig. 5. 
Two different densities determine 
two positions of axes: 
$\vec{B}_i = \vec{B}(2\theta_i)$ (i = 1,2). The angle between 
these axes,   
$
\Delta \theta \equiv 2\theta_1 - 2\theta_2~, 
$
so called the ``swing" angle, 
plays a key role in the enhancement 
mechanism. 
Let us consider an  evolution of the neutrino  
for which the  resonance condition  (\ref{pi})  is fulfilled.
Suppose, $2\theta_1 > 2\theta_2$, and  the neutrino first
propagate in the layer 1.  
This corresponds to $\vec{\nu}$ precession
around  $\vec{B}_1 = \vec{B}(2\theta_1)$.
At the border between the first  and
second layers the neutrino vector is in  position $\vec{\nu} (2)$
(which corresponds  to  phase  acquired
in the first layer,  $\Phi_1 = \pi$).
At this moment the mixing angle changes
suddenly: $\theta_1 \rightarrow \theta_2$.
In the second layer  $\vec{\nu}$   precesses around new position of
axis,  $\vec{B}_2 \equiv \vec{B}(2\theta_2)$.  
Thus after one period the neutrino turns out in a position (3) and 
the cone angle  increases as: 
$\theta_a = 2\theta_1 +   \Delta \theta$. Further, the 
cone angle will continue to increase after 
each period by the double swing angle $2\Delta \theta$.   
The cone first opens and then  shrinks in the opposite 
direction (see fig.~5).  

The enhancement depends on number of periods (perturbations) and
on the amplitude of perturbation which determines  
 the swing angle. For small perturbations,
large transition probability can be achieved after many periods.
For sufficiently large ``swing" angle even small number of periods is
enough. 

This mechanism  can be realized  
inside the Earth \cite{LS,LMS,Pet,Akh2},  
where the perturbation is large $\Delta \theta \sim 1$, 
and strong effect is achieved even after ``1.5 periods".

%%%%%%%%%%%%%%%%%%%%%%%%%%%%%%%%%%%%%%%%%%%%%%%%%%%%%%%%%%%%%%%%%%%
%%%%%%%%%%%%%%%%%%%%%%%%%  ss3.8 %%%%%%%%%%%%%%%%%%%%%%%%%%%%%%%%%%%
\subsection{Oscillations and inelastic collisions}
%%%%%%%%%%%%%%%%%%%%%%%%%%%%%%%%%%%%%%%%%%%%%%%%%%%%%%%%%%%%%%%%%%%

Another example of  significant  ($\sim O(1)$) transition  
without large  mixing angle is when the oscillations 
are accompanied by the lost of  coherence due to the 
inelastic collisions. 
At low energies the refraction length $l_0$ can be much smaller than 
the absorption length $l_{ab}$: $l_0 \ll l_{ab}$. The 
oscillation length (being of the order $l_0$) is also 
much smaller than $l_{ab}$. In this case one can consider the oscillations
between two successive inelastic collisions. Since the 
time between two collisions fluctuate, one gets averaged oscillation 
effect which is characterized by $0.5 \sin^2 2\theta_m$. 
Collision splits a neutrino state into pure flavor components and 
further  oscillations of these components will be independent. 
The process has a statistical character and the probability converges to
1/2, 
independently on mixing angle. The system approaches the 
``equillibrium". 

Graphically the effect  of absorption and 
inelastic scattering 
(a depart from the coherence)  
is equivalent to shrinking  the
neutrino vector (and the cone).  

%%%%%%%%%%%%%%%%%%%%%%%%%%%%%%%%%%%%%%%%%%%%%%%%%%%%%%%%%%%%%%%%%%
%%%%%%%%%%%%%%%%%%%%%%%%%% ss3.9 %%%%%%%%%%%%%%%%%%%%%%%%%%%%%%%%%
\subsection{Non-adiabatic perturbations}  
%%%%%%%%%%%%%%%%%%%%%%%%%%%%%%%%%%%%%%%%%%%%%%%%%%%%%%%%%%%%%%%%%%
 
Small density perturbation can lead to 
strong ``inverse" effect: 
The adiabatic transition  results in  almost 
complete transformation 
of one neutrino species into another one. 
Suppose that  the  density profile has some small perturbation 
$\Delta n$ which breaks adiabaticity. If a  size  of the  perturbation 
is comparable with the size of the resonance layer: 
\begin{equation} 
\frac{\Delta n}{n} \sim \sin 2\theta~, 
\end{equation} 
the  perturbation will induce  transition
with $P \sim 1$ to the 
original neutrino flavor in certain  energy range \cite{KS}. 
%That is, one expects the appearance of the peak in transition 

%%%%%%%%%%%%%%%%%%%%%%%%%% s4  %%%%%%%%%%%%%%%%%%%%%%%%%%%%%%%%%%
%%%%%%%%%%%%%%%%%%%%%%%%%%%%%%%%%%%%%%%%%%%%%%%%%%%%%%%%%%%%%%%%%%
\section{Dynamics in the multilevel system}
%%%%%%%%%%%%%%%%%%%%%%%%%%%%%%%%%%%%%%%%%%%%%%%%%%%%%%%%%%%%%%%%%%

Dynamics of transitions in a system of three (or  
more) neutrino species is, of course, much more complicated 
than in the 2$\nu-$ case.   
New effects appear, {\it  e.g.}, CP-violation. 
%We will consider some elements of this dynamics. 
In certain realistic situations ( mass hierarchy, smallness of mixing)  
the task can be reduced to the evolution 
in two neutrino systems. There are also some generic 
$3\nu-$ effects which exist  even in the case of mass hierarchy.

Let us consider the  neutrino mass spectrum with: 
\begin{equation}
\Delta m^2_{12}\ll \Delta m^2_{23} \approx \Delta m^2_{13}~.  
\end{equation}
The dynamics of propagation can be reduced to the $2\nu$- case  
in the following circumstances. 

%%%%%%%%%%%%%%%%%%%%%%%%%%%%%%%%%%%%%%%%%%%%%%%%%%%%%%%%%%%%%%%
%%%%%%%%%%%%%%%%%%%% ss4.1 %%%%%%%%%%%%%%%%%%%%%%%%%%%%%%%%%%%%
\subsection{Short range experiment. Freeze out of  subsystem.}  
%%%%%%%%%%%%%%%%%%%%%%%%%%%%%%%%%%%%%%%%%%%%%%%%%%%%%%%%%%%%%%%

Suppose the source - detector distance $d$ is much smaller than the 
oscillation length associated with  smallest mass splitting:   
\begin{equation} 
d \ll l_{\nu}  = \frac{4 \pi E}{\Delta m^2_{12}}~. 
\end{equation}
In this case the phase difference between 
$\nu_1$ and $\nu_2$ acquired on the way $d$ will be very small: 
$\phi_{12} = 2\pi d/l_{\nu} \ll 1$, 
which means that  there is no evolution 
in the subsystem $\nu_1 - \nu_2$. 
This subsystem is ``frozen". 

Let us consider, for instance, the decomposition of  the $\nu_e$:  
\be 
\nu_e = U_{e1} \nu_1 +  U_{e2} \nu_2  +  U_{e3}\nu_3. 
\label{nue}
\ee 
Since the internal evolution in the $\nu_1 -  \nu_2$ subsystem is frozen, 
we can consider this subsystem as the  unique state 
\be
\tilde{\nu} \equiv 
\cos \theta_{12} \nu_1 +  \sin \theta_{12}  \nu_2~,     
\ee
where 
\be
\cos \theta_{12}  = \frac{U_{e2}}{\sqrt{ U_{e1}^2 + U_{e2}^2}} 
\equiv \frac{U_{e2}}{\sqrt{1 - U_{e3}^2}}~. 
\ee
Now $\nu_e$ state can be rewritten as 
\be
\nu_e = \sqrt{1 - U_{e3}^2}~\tilde{\nu}  +  U_{e3}~\nu_3~, 
\ee
and  the task is reduced to  evolution of the 
$2\nu$- system $\tilde{\nu} -  \nu_3$ with the  
effective mixing parameter $\sin \theta = U_{e3}$  and 
$\Delta m^2 = m_3^2$. This is the so called  one level 
dominating scheme,  when the oscillations are determined by 
flavor composition and the mass of the heaviest state $\nu_3$ \cite{one}. 

%%%%%%%%%%%%%%%%%%%%%%%%%%%%%%%%%%%%%%%%%%%%%%%%%%%%%%%%%%%%%%%%%%%%%%%
%%%%%%%%%%%%%%%%%%%%%%%%%%   ss4.2 %%%%%%%%%%%%%%%%%%%%%%%%%%%%%%%%%%%%%
\subsection{Long range experiments. Decoupling of one state.} 
%%%%%%%%%%%%%%%%%%%%%%%%%%%%%%%%%%%%%%%%%%%%%%%%%%%%%%%%%%%%%%%%%%%%%%

Let us consider the case when the source-detector distance 
is much larger than the oscillation length associated with  
the  largest mass splitting:  
\begin{equation}
d \gg l_{\nu}  = \frac{4 \pi E}{\Delta m^2_{23}}.
\end{equation}
Oscillations due to $\Delta m^2_{23}$ 
are usually averaged out or/and the coherence of the $\nu_3$ 
with the rest of system is lost. The state  $\nu_3$ 
decouples, leading to the averaged oscillation result. 
Nontrivial evolution will be in the $\nu_1 -  \nu_2$ subsystem. 
Let us consider again the propagation of the $\nu_e$ (\ref{nue}). 
Taking into account  decoupling of the $\nu_3$ we can write 
immediately  the survival probability as \cite{dec}
\be
P = (1 - U_{e3}^2)^2  P(\Delta m_{12}^2, \theta)   +  U_{e3}^4~.
\ee
Here $P(\Delta m_{12}^2, \theta)$ is the survival  
probability in  
$\tilde{\nu} \leftrightarrow \tilde{\nu}'$ transition,  where 
$\tilde{\nu}' = \cos \theta_{12} \nu_2 -  \sin \theta_{12}  \nu_1$ 
is the state orthogonal to $\tilde{\nu}$.

Notice that for solar neutrinos both regimes can be realized 
if $\Delta m_{12}^2 \sim 10^{-10}$ eV$^2$ and 
$\Delta m_{13}^2 \sim 10^{-5}$ eV$^2$. Indeed,    
on the way from the center of the sun to its surface  the 
$\nu_1 - \nu_2$ subsystem is frozen, whereas on the 
way from the surface of the Sun to the Earth the state $\nu_3$
decouples.  

%%%%%%%%%%%%%%%%%%%%%%%%%%%%%%%%%%%%%%%%%%%%%%%%%%%%%%%%%%%%%%%%%%%%%%
%%%%%%%%%%%%%%%%%%%%%%%%%%%%   ss4.3 %%%%%%%%%%%%%%%%%%%%%%%%%%%%%%%%%
\subsection{Generic $3 \nu$-effect} 
%%%%%%%%%%%%%%%%%%%%%%%%%%%%%%%%%%%%%%%%%%%%%%%%%%%%%%%%%%%%%%%%%%%%%%%

In matter in the $3\nu$-case, the subsystem 
$\nu_1 - \nu_2$ can give significant 
oscillation effect even 
for  very small vacuum splitting $\Delta m_{12}^2$   
\cite{pant,giunti,fogli}. Indeed,  matter gives the  
contribution to the level splitting,  
$\Delta H \approx V$,  which dominates in the case 
$\Delta m_{12}^2/2E \ll V$. Therefore 
even for small $\Delta m_{12}^2$,  the splitting, and consequently, the  
phase of oscillations can be large. However, the oscillation effect 
will be still strongly 
suppressed in the 2$\nu$-case since  with increase of splitting the
effective mixing decreases: $\theta_m \propto \Delta m^2_{12} / 2 E V$.  
As the consequence,  the  oscillations will have very small depth. 

Such a situation can be avoided in schemes with three neutrino mixing  
and significant admixture of  $\nu_e$ in  the heaviest state 
$\nu_3$. Suppose  the potential 
satisfies  inequality: 
\be
\frac{m_{2}^2}{2E} \ll V \ll \frac{m_{3}^2}{2E}~.  
\label{condi}
\ee
The key point is that at  this condition the matter 
does not change practically the flavors of the $\nu_3$.  
In particular,  the admixture of $\nu_e$,  $U_{e3}$, 
will be unsuppressed. 
At the same time  matter changes strongly  flavors of 
two other eigenstates. 

Let us consider the $\nu_{\mu} \leftrightarrow \nu_{\tau}$ oscillations 
due to mixing of two lightest eigenstates 
1 - 2  with splitting  $\Delta H_{12} \approx V$. 
The mixing of these eigenstates would be absent,  if 
$\nu_{2m}$ had pure electron neutrino flavor. 
This occurs in the 2$\nu$ scheme. However, in the $3\nu$-case, a  
part of the $\nu_e$-flavor is in  $\nu_3$. Therefore
due to unitarity the admixture of $\nu_e$ in $\nu_{2m}$  should be 
smaller than 1:   
\be 
U_{e 2}^m =  \sqrt{1 - U_{e3}^2} <1~, 
\ee
and correspondingly, the admixtures    of 
$\nu_{\mu}$ and $\nu_{\tau}$ flavors in $\nu_{2m}$ should not vanish. 
As a  consequence, light states are mixed and the  
$\nu_{\mu} \leftrightarrow \nu_{\tau}$   oscillations exist  
with unsuppressed depth. 
It is easy to find flavors of the neutrino eigenstates 
in medium $U_{f i}^m$ at the conditions (\ref{condi}): 
\begin{equation}
\begin{array}{ll}
U_{\mu 2}^m = (U_{e 2}^m)^{-1} U_{e 3} U_{\mu 3}, &
U_{\tau 2}^m = (U_{e 2}^m)^{-1} U_{e 3} U_{\tau 3},\\
U_{\mu 1}^m = (U_{e 2}^m)^{-1}  U_{\tau 3}, & 
U_{\tau 1}^m = (U_{e 2}^m)^{-1} U_{\mu 3}. 
\end{array} 
\end{equation}
From here we get  the depth of $\nu_{\mu} \leftrightarrow \nu_{\tau}$   
oscillations  
\be
4 U_{\mu 1}^m U_{\tau 1}^m U_{\mu 2}^m U_{\tau 2}^m = 
\frac{4U_{e 3}^2 U_{\mu 3}^2 U_{\tau 3}^2}{(1 - U_{e3}^2)^2} 
\ee
which does not depend on matter density.  
Clearly,  the oscillations disappear if $U_{e 3} = 0$.  
Such an effect can be relevant for the atmospheric neutrinos.\\ 

%%%%%%%%%%%%%%%%%%%%%%%%%% s.5 %%%%%%%%%%%%%%%%%%%%%%%%%%%%%%%%%%%%%
%%%%%%%%%%%%%%%%%%%%%%%%%%%%%%%%%%%%%%%%%%%%%%%%%%%%%%%%%%%%%%%%%%%%
\section{Neutrino spectra and Neutrino Transitions}
%%%%%%%%%%%%%%%%%%%%%%%%%%%%%%%%%%%%%%%%%%%%%%%%%%%%%%%%%%%%%%%%%%%%

Let us consider  phenomenology 
of different neutrino mass schemes. We will  
assume that all neutrino masses are 
below a few  eV,   so that both cosmological and structure formation 
bounds are satisfied without neutrino decay.  
We will concentrate on applications of the neutrino  transitions 
considered in sect. 3  
to  supernova neutrinos \cite{sn,snafter}.

In supernova the density changes on the way of of neutrinos 
from nuclear values to  practically zero. 
With a such  profile  one can probe 
whole  mass spectrum of neutrinos. Indeed, 
the resonance density can be estimated as 
\be 
\rho_R \sim 10^{6} {\rm g/cm}^{3}~\frac{\Delta m^2 }{1 {\rm eV}^2}~.  
\label{resd}
\ee
Since $\Delta m^2 \leq m^2 <$ few eV$^2$ , all the transitions 
(with one possible exception) will occur far outside the core   
which means that they do not influence collapse. 
The transitions can, however, influence the nucleosynthesis 
in the internal parts of star \cite{r-proc}
\be 
\rho_{NS} = (10^{6} - 10^{10})~ {\rm g/cm}^{3}~. 
\ee

The transitions change properties of fluxes observed at the Earth. 
They occur in the resonance layers as well as on the 
way from the star to the Earth. 
The efficiency of transition in a given resonance is determined  by the 
adiabaticity.  The edge of the 
adiabatic domain in the ($\Delta m^2 - \sin^2 2\theta$)- 
plot can be described roughly by 
\be 
\sin^2 2\theta > 
10^{- 5}\left(\frac{\Delta m^2}{1 {\rm eV}^2}\right)^{-3/4}~, 
\label{adiab}
\ee
where $\theta$ is the vacuum mixing  angle of the resonating states.  
In the adiabatic domain the survival  probability equals 
$P_{\alpha - \alpha} \approx \sin^2 \theta$. 

Since $\nu_{\mu}$ and  $\nu_{\tau}$ have identical 
production and detection properties, 
we can consider phenomenology in terms of any combinations of these
states. We will denote these neutrinos as the non-electron,  
$\nu_{ne}$, neutrinos.  

%%%%%%%%%%%%%%%%%%%%%%%%%%%%%%%%%%%%%%%%%%%%%%%%%%%%%%%%%%%%%%%%%%%%
\subsection{Effects of neutrino conversion}
%%%%%%%%%%%%%%%%%%%%%%%%%%%%%%%%%%%%%%%%%%%%%%%%%%%%%%%%%%%%%%%%%%%%%

The neutrino transformations  in supernovae  lead 
to the following effects.  

\noindent
1. Disappearance of the neutronization peak. In the 
case of strong $\nu_{e} \rightarrow \nu_s$ conversion the peak 
will not be observable both  in the charged and  in the  neutral current
interactions. 

\noindent
2. Change of flavor of the neutronization peak. The 
oscillations/conversion $\nu_{e} \rightarrow \nu_{ne}$, where 
$\nu_{ne} \equiv  \nu_{\mu},  \nu_{\tau}$,   lead  to   
appearance of the $\nu_{ne}$-neutronization peak. 
This effect can be detected by comparison of signals 
in the charged current (sensitive to $\nu_{e}$ only) 
and neutral current (sensitive to whole flux) 
interactions. 

\noindent
3. Modification of spectrum of the electron neutrinos 
during cooling stage. The $\nu_{e}$-spectrum at the Earth is 
\be
F_e (E) = F_e^0 (E) P_{e \rightarrow e}(E) + 
F_x^0 (E) P_{x \rightarrow e}(E), 
\ee
where $F_e^0$ and $ F_x^0$ are  
the  original spectra  of the
electron neutrinos and non-electron ($\nu_{\mu}$,  $\nu_{\tau}$)  
neutrinos respectively. 
$F_e^0$ is the soft and 
$F_x^0$ is the hard component:  
the average energies of spectra satisfy  inequality  
$E(\nu_{e}) < E(\nu_{x})$ 
and the difference follows from  the difference of 
interactions of these neutrinos. 
$P_{e \rightarrow e}$ 
and  
$P_{x \rightarrow e}$ 
are the conversion probabilities which may not be 
related in the multilevel system:  
$P_{x \rightarrow e} \neq 1 - P_{e \rightarrow e}$.

If $P_{e \rightarrow e} \approx 0$ 
and 
$P_{x \rightarrow e}\approx 1$, 
there is a complete interchange of the spectra. For the 
energy independent 
probabilities in the $2\nu-$ case  the effects can be  characterized 
by  ``permutation" parameter \cite{smi}: 
$p \equiv P_{x \rightarrow e}$, 
$P_{e \rightarrow e} \equiv  1 - p$.   
Complete interchange of spectra corresponds to 
$p = 1$. For $p < 1 $ (partial permutation),  
$F_e (E)$ will have both soft and hard components. 

If  $P$ depends on energy  
the distortion of the spectrum can be more complicated.

\noindent
4. Modification of the $\bar{\nu}_{e}$ spectrum. 
It can be desribed in a similar way. 

Notice that transition  effects in the neutrino 
and antineutrino channels are usually different. 
The resonance transitions 
(due to mass) in the neutrino channels 
are not accompanied by the transitions in the 
antineutrino channels.

Depending on the level crossing scheme one 
predicts different combinations 
of the above effects. In what follows we will 
consider predictions from different
schemes and discuss the possibility to identify 
the scheme \cite{DS}.

%%%%%%%%%%%%%%%%%%%%%%%%%%%%%%%%%%%%%%%%%%%%%%%%%%%%%%%%%%%%%%%%%%%%%%
%%%%%%%%%%%%%%%%%%%%%%%%  ss5.2     %%%%%%%%%%%%%%%%%%%%%%%%%%%%%%%%%%%
\subsection{$3\nu$ scheme for solar and atmospheric neutrinos}
%%%%%%%%%%%%%%%%%%%%%%%%%%%%%%%%%%%%%%%%%%%%%%%%%%%%%%%%%%%%%%%%%%%%%%%

Let us consider the hierarchical mass spectrum with 
\be
m_3 = (0.3 - 1) \cdot 10^{-1} {\rm eV},~~~
m_2 = (2 - 4) \cdot 10^{-3}{\rm eV},~~~
m_1 \ll  m_2
\ee
(see fig. 6). $\nu_{\mu}$ and $\nu_{\tau}$ mix strongly 
in $\nu_{2}$ and $\nu_{3}$. The electron flavor 
is weakly mixed: 
it  is mainly in $\nu_{1}$ with small admixtures 
in the heavy states. Such a scheme has the following properties: 

%%%%%%%%%%%%%%%%%%%%%%%%%%%%%%%%%%%%%%%%%%%%%%%%%%%%%%%%%%%%%%%%%%%%%%

%%%%%%%%%%%%%%%%%%%%%%%%%%%%%%%%%%%%%%%%%%%%%%%%%%%%%%%%%%%%%%%%%%%%%%
%%%%%%%%%%%%%%%%%%%%%%%%% stan.eps %%%%%%%%%%%%%%%%%%%%%%%%%%%%%%%%%%%%%
\begin{figure}[htb]
\hbox to \hsize{\hfil\epsfxsize=8cm\epsfbox{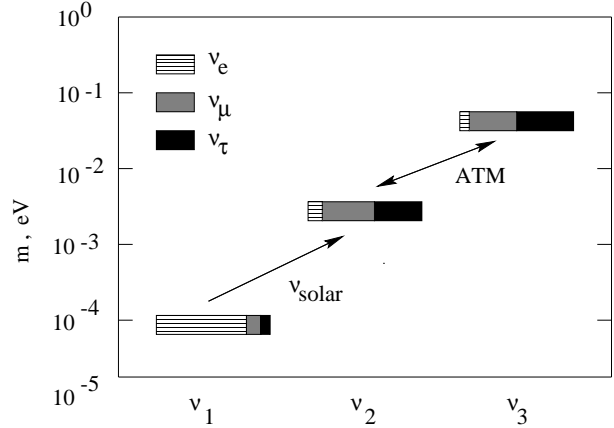}\hfil}
\caption{~~Neutrino mass and mixing 
pattern  of  the scheme for the  solar and atmospheric
neutrinos. The boxes correspond to the mass eigenstates. The sizes 
of different regions in the boxes show admixtures 
of different flavors. 
Weakly hatched regions correspond to the electron 
flavor, strongly hatched regions  depict the muon flavor, black regions 
present the tau flavor.   
}
\end{figure}
%%%%%%%%%%%%%%%%%%%%%%%%%%%%%%%%%%%%%%%%%%%%%%%%%%%%%%%%%%%%%%%%%%%%%%
\noindent
(i) It  explains 
the solar neutrino data via $\nu_e \rightarrow \nu_{2}$ 
resonance conversion inside the Sun. Notice 
that $\nu_e$ converts to $\nu_{\mu}$ and $\nu_{\tau}$ in  comparable
portions. 

\noindent
(ii) The atmospheric neutrino anomaly is solved via 
$\nu_{\mu} \leftrightarrow \nu_{\tau}$ oscillations. Small $\nu_e$
admixture 
in $\nu_{3}$ can lead to resonantly 
enhanced oscillations in  matter of
the Earth. 

\noindent
(iii) There is no explanation of  the LSND result,  and 
the contribution to the Hot Dark Matter component of the universe is 
small:  $\Omega_{\nu} < 0.01$.   

The scheme can be probed by  the long baseline experiments. 

%%%%%%%%%%%%%%%%%%%%%%%%%%%%%%%%%%%%%%%%%%%%%%%%%%%%%%%%%%%%%%%%%%%
%%%%%%%%%%%%%%%%%%%%%%%  generic  %%%%%%%%%%%%%%%%%%%%%%%%%%%%%%%%%
\begin{figure}[htb]
\hbox to \hsize{\hfil\epsfxsize=7cm\epsfbox{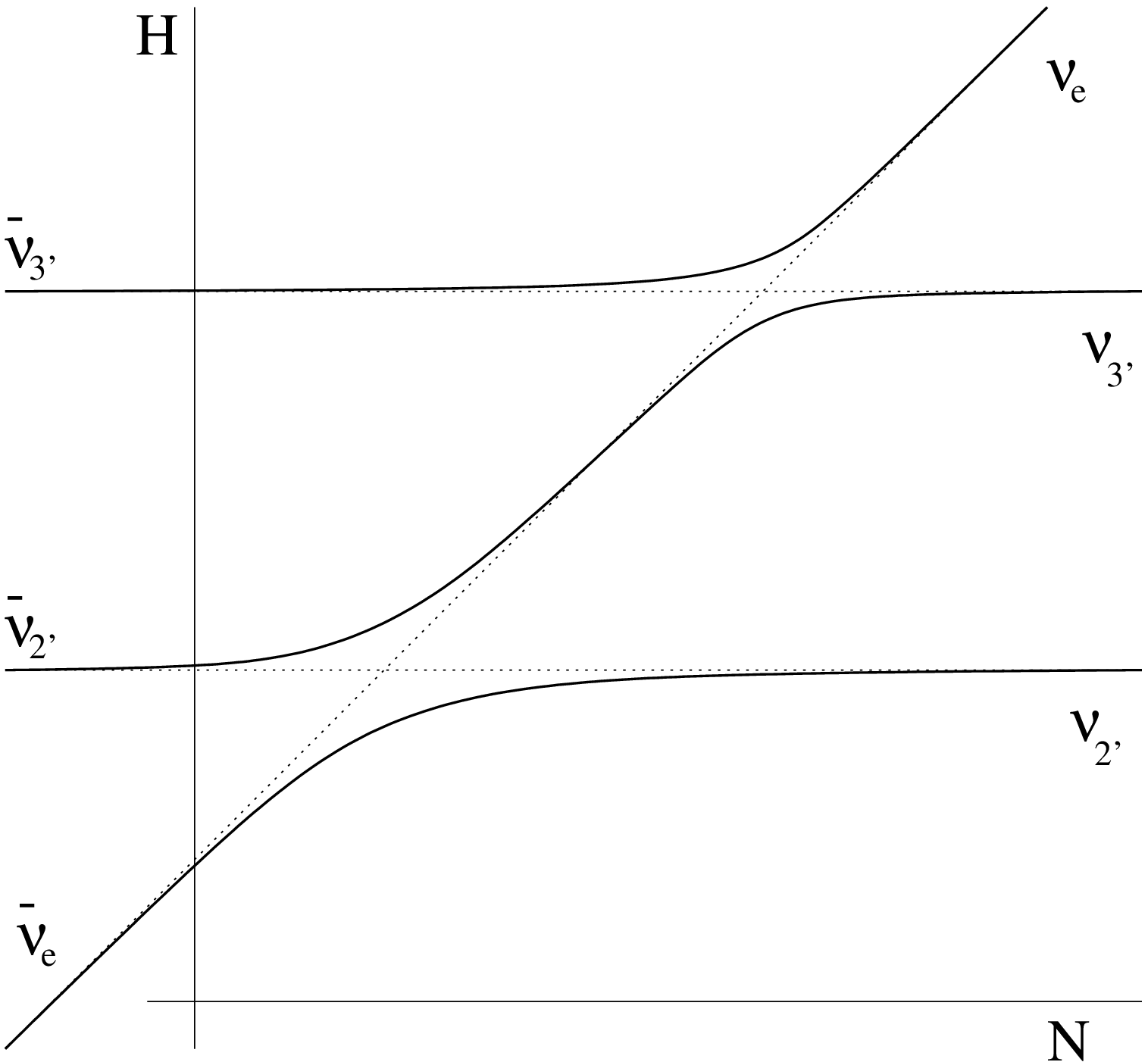}\hfil}
\caption{~~The level crossing pattern  of 
the scheme for the solar and atmospheric neutrinos. 
Solid lines show the eigenvalues of the system as functions of the
density. The dashed lines correspond to energies of 
$\nu_e$, $\nu_2'$, and  $\nu_3'$.   
The part of the plot with $N < 0$  corresponds to the 
antineutrino channels. 
}
\end{figure}
%%%%%%%%%%%%%%%%%%%%%%%%%%%%%%%%%%%%%%%%%%%%%%%%%%%%%%%%%%%%%%%%%%%  
It is convenient to consider neutrino transitions in the basis 
$\nu_e$, $\nu_2'$, $\nu_3'$, where  
$\nu_2'$ and  $\nu_3'$  diagonalize the 
mass  matrix for $\nu_{\mu} - \nu_{\tau}$ 
subsystem.  
The mixing of $\nu_e$ with  $\nu_2'$  and $\nu_3'$ is small: 
Since $\nu_2'$  and $\nu_3'$ coincide up to small 
corrections with mass eigenstates this mixing is determined 
by $U_{ei}$, $i = 2, 3$.  
$U_{e2}$  is fixed by the solar neutrino data,  and 
$U_{e3}$ is  weakly restricted by the atmospheric neutrino results
and reactor bounds. 

In fig.~7 we show the level crossing scheme:  
the dependence of the eigenvalues on the density of medium. 
If $U_{e2}$ is large enough,  then all the level crossings are adiabatic,   
and the following transitions occur in a supernova: 
\be
\nu_e \rightarrow \nu_3',~~~
\nu_3' \rightarrow \nu_2',~~~
\nu_2' \rightarrow \nu_e.  
\ee
All these transitions will occur in the outer layers of the 
stars, and  therefore they do  not influence 
collapse and nucleosynthesis. The transitions, however,  modify  
fluxes expected at the Earth.  
The neutronization $\nu_e$ peak disappears. 
Instead one would expect $\nu_{\mu}/\nu_{\tau}$ neutronization peak 
which can be detected by the neutral current interactions. 
The $\nu_e$ from cooling stage will have the hard spectrum $F^0_x$, 
the spectrum of non-electron neutrinos, 
$\nu_{\mu}$ and $\nu_{\tau}$, will 
contain both the  soft (original $\nu_e$) and the hard components. 
The antineutrino signal is unchanged. 
Similar modifications are expected  if one (among two)  resonance 
crossings is non-adiabatic. 

The modification of the scheme 
is possible in which all three neutrinos  have approximately 
the same mass $m_0$ (almost degenerate)  
but with  $\Delta m^2$ and mixings as before. In this case neutrinos 
give  a significant ($\Omega_{\nu} \sim 0.1$) contribution to the HDM. 
Since the $\nu_e$ dominates in one of the mass 
eigenstates, the effective Majorana mass relevant for 
the neutrinoless double beta decay is about $m_0$ and  searches 
of the $\beta \beta_{0\nu}$ decay give crucial check of this 
version \cite{MY}. 

%%%%%%%%%%%%%%%%%%%%%%%%%%%%%%% ss4.3 %%%%%%%%%%%%%%%%%%%%%%%%%%%%%%%%
%%%%%%%%%%%%%%%%%%%%%%%%%%%%%%%%%%%%%%%%%%%%%%%%%%%%%%%%%%%%%%%%%%%%%%
\subsection{Bi-maximal  and bi-large mixings} 
%%%%%%%%%%%%%%%%%%%%%%%%%%%%%%%%%%%%%%%%%%%%%%%%%%%%%%%%%%%%%%%%%%%%%

{}The SK data on atmospheric neutrinos give   
strong evidence that mixing in the 
$\nu_{\mu} - \nu_{\tau}$ channel is large (or even maximal). 
Probably  mixing is large in   other  channels.  
In this context several schemes where elaborated 
\cite{bimax}. 

In the bi-maximal scheme  neutrinos have masses 
\be
m_3 = (0.3 - 3) \cdot 10^{-1} {\rm eV},~~~
m_2 \sim  10^{-5}{\rm eV},~~~
m_1 \ll  m_2
\ee
(see fig. 8). 
$\nu_{\mu}$ and $\nu_{\tau}$ mix maximally in  
$\nu_{3} = (\nu_{\mu} + \nu_{\tau})/\sqrt{2}$. The orthogonal
combination, $\nu_2' \equiv (\nu_{\mu} - \nu_{\tau})/\sqrt{2}$  
strongly mixes with 
$\nu_e$ in $\nu_{1}$  and  $\nu_{2}$. 
There is no admixture of $\nu_e$ in the $\nu_{3}$.  
%%%%%%%%%%%%%%%%%%%%%%%%%%%%%%%%%%%%%%%%%%%%%%%%%%%%%%%%%%%%%%%%%
%%%%%%%%%%%%%%%%%%%%%%%%%%%% bima  %%%%%%%%%%%%%%%%%%%%%%%%%%%%%%
\begin{figure}[htb]
\hbox to \hsize{\hfil\epsfxsize=8cm\epsfbox{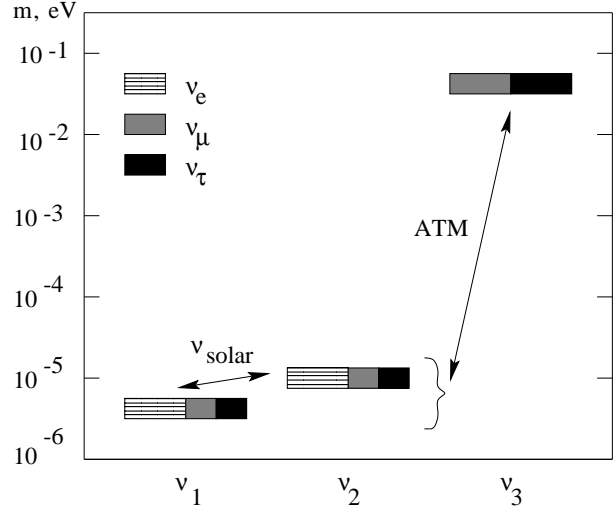}\hfil}
\caption{~~The neutrino mass and mixing pattern  of the  bi-maximal
mixing  scheme.  
}
\end{figure}
%%%%%%%%%%%%%%%%%%%%%%%%%%%%%%%%%%%%%%%%%%%%%%%%%%%%%%%%%%%%%%%%%
In this scheme 

\noindent
(i)  The solar neutrino problem  can be solved via 
$\nu_e \leftrightarrow \nu_2'$  ``Just-so" vacuum oscillations. 
Notice 
that $\nu_e$ converts equally to $\nu_{\mu}$ and $\nu_{\tau}$. 

\noindent
(ii) The atmospheric neutrino anomaly is solved via 
$\nu_{\mu} \leftrightarrow \nu_{\tau}$ maximal depth oscillations. 

The spectrum can supply significant amount of the HDM if all three
neutrinos are strongly degenerate. 

In fig.~9 we show the level crossing scheme of  the spectrum 
in  general case when there is small admixture 
of the $\nu_e$ state in $\nu_3$. This can be checked 
by searches for an  excess of the e-like events 
in the atmospheric neutrinos.  
In the strict  bi-maximal mixing case, when $U_{e3} = 0$, 
the state $\nu_3$ decouples and 
$\nu_e$ mixes maximally with $\nu_2'$.   
%%%%%%%%%%%%%%%%%%%%%%%%%%%%%%%%%%%%%%%%%%%%%%%%%%%%%%%%%%%%%%%%%%%%%
%%%%%%%%%%%%%%%%%%%% almost-bimax %%%%%%%%%%%%%%%%%%%%%%%%%%%%%%%%%%%
\begin{figure}[htb]
\hbox to \hsize{\hfil\epsfxsize=7cm\epsfbox{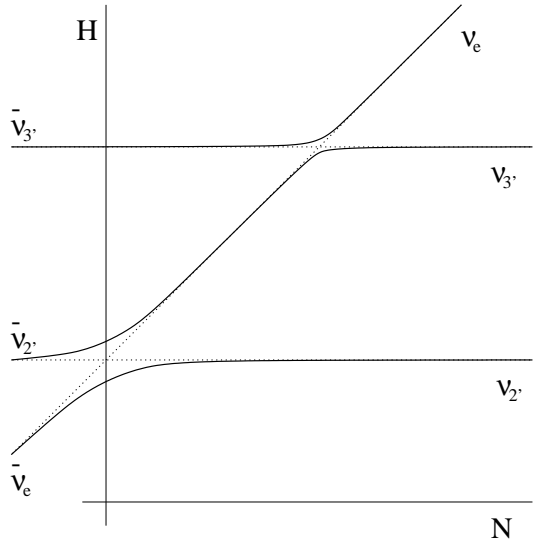}\hfil}
\caption{~~The level crossing pattern of the  bi-maximal mixing  
scheme with small admixture of the $\nu_e$ in the 
heavy state. 
}
\end{figure}
%%%%%%%%%%%%%%%%%%%%%%%%%%%%%%%%%%%%%%%%%%%%%%%%%%%%%%%%%%%%%%%%%%%%%

For supernova neutrinos we predict the following. 
The  electron neutrinos  
oscillate into combination of $\nu_{\mu}$ and $\nu_{\tau}$, 
with maximal depth on the way from 
collapsing star to the Earth. 
Similarly,  the  electron antineutrinos  
oscillate into $\bar{\nu}_{\mu}$ and $\bar{\nu}_{\tau}$.  
As the result the neutronization peak will consist of 
equal number of the electron and non-electron 
neutrinos (which could be checked by comparison of 
signals due to   the neutral and charged currents).

Also the spectra from the cooling stage will be modified.  
In particular, the $\bar{\nu}_e$-spectrum  will have both  
the soft (original $\nu_e$) component 
and  the hard component (original $\nu_{\mu}$) in equal portions:   
the permutation parameter is 0.5. 
The same holds for ${\nu}_e$. 

Situation can  be different,  
if there is some admixture of the $\nu_e$ 
in $\nu_3$. Now the $\nu_e - \nu_3'$ level crossing  occurs
 (fig.~9), and if the
adiabaticity condition is fulfilled one expects: 
\be
\nu_e \rightarrow \nu_3',~~~
\nu_3' \rightarrow \nu_2',~~~
\nu_2' \rightarrow \nu_1.
\ee
Recall that $\nu_1$ is the maximal mixture of the electron and
non-electron
neutrinos.  
Therefore one expects: 
(i) complete (in contrast with previous case) disappearance of the 
$\nu_e$ neutronization peak and 
appearance of the peak of  non-electron neutrinos;  
(ii) the electron neutrinos with the  hard spectrum 
(of original $\nu_{\mu}$);  
(iii) muon and tau neutrinos with  both the  soft 
and the hard components.   
(iv) At the same time  $\bar{\nu}_e$ will have composite spectrum 
with hard and soft components. This distinguishes  bi-maximal 
scheme from that of sect. 5.2.  

The  mixing of ${\nu_e}$ can be  
non-maximal but large. If then $m_2 \sim (3 - 4) \times 10^{-3}$  eV,  
the solar neutrino deficit can be explained by  
the large mixing angle MSW solution. 
The consequences for supernova neutrinos are rather similar 
to previous case. At the same time  the permutation parameter for 
$\bar{\nu}_e$ can be smaller. That is,  the contribution of 
the hard component to $\bar{\nu}_e$ spectrum is smaller. 

One can introduce a  degeneracy of neutrinos
(keeping the same $\Delta m^2$) to get significant amount the HDM 
in the Universe. Now  the effective 
Majorana mass of the electron neutrino can be 
small due to cancellation  related to large mixing. 

Let us comment on the version of the 
bi-maximal scheme with inverted mass
hierarchy:  
$
m_1 \approx m_2 \gg m_3 ~, 
$
when
two states with maximal (or large) $\nu_e$ mixing 
are heavy and degenerate, 
whereas the third state  with large   
$\nu_{\mu} - \nu_{\tau}$ mixing and small $\nu_e$ admixture  is light. 
In this scheme the $\nu_e - \nu_3'$ level crossing 
occurs in the {\it antineutrino} semiplane, so that 
in the supernova  $\bar{\nu}_e$ will be strongly converted into 
combination of $\bar{\nu}_{\mu} - \bar{\nu}_{\tau}$ and vice versa.
As the result the $\bar{\nu}_e$'s will have hard spectrum.  

%%%%%%%%%%%%%%%%%%%%%%%%%%% ss5.4  %%%%%%%%%%%%%%%%%%%%%%%%%%%%%%%%%%
%%%%%%%%%%%%%%%%%%%%%%%%%%%%%%%%%%%%%%%%%%%%%%%%%%%%%%%%%%%%%%%%%%%%%
\subsection{Models with sterile neutrinos} 
%%%%%%%%%%%%%%%%%%%%%%%%%%%%%%%%%%%%%%%%%%%%%%%%%%%%%%%%%%%%%%%%%%%%%%

There are two motivations for the introduction of 
sterile neutrinos:  
(i) to reconcile different neutrino anomalies 
including  the LSND result;  
(ii) to explain an existence of the large mixing in the 
leptonic sector  (in contrast with quark sector). 
Large mixing  implied by  the atmospheric neutrino 
data can be  the mixing of  $\nu_{\mu}$  with sterile neutrino. All  
flavor mixings can be small. 
There is another indirect connection related to the fact that  
large (maximal) mixing prefers degeneracy of mass (see sect. 5.6).

If the atmospheric neutrino problem is solved due to oscillations of 
$\nu_{\mu}$ and $\nu_{\tau}$ strongly mixed 
in degenerate states, then there is no way to solve the solar neutrino
problem. For this one can introduce sterile neutrino 
which mixes with $\nu_e$.

%%%%%%%%%%%%%%%%%%%%%%%%%%%  ss5.5 %%%%%%%%%%%%%%%%%%%%%%%%%%%%%%%%%%%%  
%%%%%%%%%%%%%%%%%%%%%%%%%%%%%%%%%%%%%%%%%%%%%%%%%%%%%%%%%%%%%%%%%%%%%%%%
\subsection{Intermediate mass scale scenario}

Intermediate mass scale scenario is characterized by neutrino mass
hierarchy, small mixing,   
and the Majorana masses of the  right handed neutrinos 
(in the context of the see-saw) at the intermediate mass scale: 
$10^{10} - 10^{13}$ GeV. In addition, the light singlet fermion 
can be  introduced  to solve the atmospheric neutrino problem 
\cite{LS} (fig.~10).  
%%%%%%%%%%%%%%%%%%%%%%%%%%%%%%%%%%%%%%%%%%%%%%%%%%%%%%%%%%%%%%%%%%%%%
%%%%%%%%%%%%%%%%%%%%%%%%%%%%%%%%% int.eps %%%%%%%%%%%%%%%%%%%%%%%%%%%
\begin{figure}[htb]
\hbox to \hsize{\hfil\epsfxsize=9cm\epsfbox{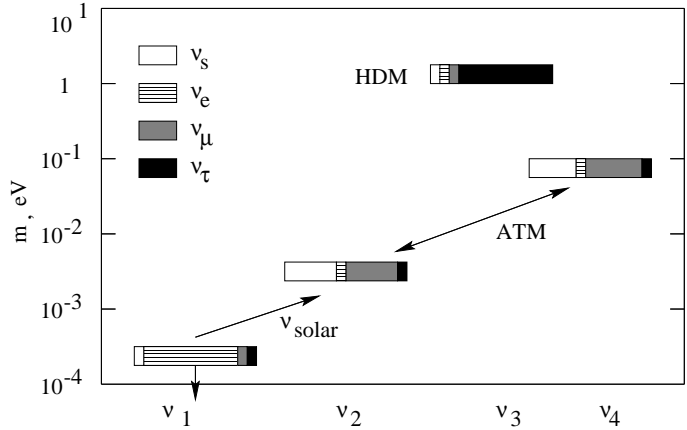}\hfil}
\caption{~~Pattern of the neutrino mass and mixing in the 
intermediate mass scale scenario. Here white parts of boxes 
correspond to the sterile state. 
}
\end{figure}
%%%%%%%%%%%%%%%%%%%%%%%%%%%%%%%%%%%%%%%%%%%%%%%%%%%%%%%%%%%%%%%%%%%%%
The  neutrino masses equal  
\be
m_4 = (0.3 - 3) \cdot 10^{-1} {\rm eV},~~~
m_2 \sim 3 \times 10^{-3}{\rm eV},~~~
m_3 \sim 1 {\rm eV}, ~~~
m_1 \ll  m_2.  
\ee 
In this scheme 
$\nu_s$ and $\nu_{\mu}$ are strongly mixed in the $\nu_2$ and $\nu_4$
eigenstates, 
so that  $\nu_{\mu} \leftrightarrow  \nu_s$ oscillations 
solve the  atmospheric neutrino problem;  
$\nu_e \rightarrow \nu_{\mu}, \nu_{s}$ resonance conversion  
explains the solar neutrino data, and $\nu_3$ can 
supply significant amount of the HDM. 

The level crossing pattern (fig. 11) can be constructed in the following
way. Let us introduce the  eigenstates of  
strongly mixed subsystem $\nu_{\mu}-  \nu_s$: 
$\nu_2'$ and $\nu_4'$. At zero density they coincide 
with $\nu_2$ and $\nu_4$ up to small  
admixitures of the $\nu_e$ and $\nu_{\tau}$. 
In the basis  of the states ($\nu_e, \nu_2', \nu_4', \nu_{\tau}$)
 all the mixings are small.  As follows from fig.~11,    
there are  four resonances in the neutrino channels  and no level
crossing in the antineutrino channels. 

Let us consider the effects in  supernova \cite{LS}. 
The transitions   can be important for the
nucleosynthesis of heavy elements due to $r$-processes.  
Above the  $\nu_e - \nu_{\tau}$ resonance at 
$\rho > 10^{8}(m_3/5 eV)^2$ g/cm$^{3}$ there is unchanged 
$\nu_e$-flux, whereas at 
$
\rho < 10^{8}(m_3/5 {\rm eV})^2    {\rm g/cm}^{3} 
$
the $\nu_e$- flux disappears thus producing conditions for 
neutron reach medium desired for r-processes.  
Indeed, in the $\nu_e - \nu_{\tau}$ resonance,  $\nu_e$  are
transformed to $\nu_{\tau}$, however inverse transition is absent: 
$\nu_{\tau}$ are converted to $\nu_s$ at about two times 
larger density and there is no inverse transition 
since there is no original flux of sterile neutrinos \cite{peltsn} 
(see however sect. 5.8). 
%%%%%%%%%%%%%%%%%%%%%%%%%%%%%%%%%%%%%%%%%%%%%%%%%%%%%%%%%%%%%%%%%%
%%%%%%%%%%%%%%%%%%%%%%%  intermediate.eps %%%%%%%%%%%%%%%%%%%%%%%%
\begin{figure}[htb]
\hbox to \hsize{\hfil\epsfxsize=8cm\epsfbox{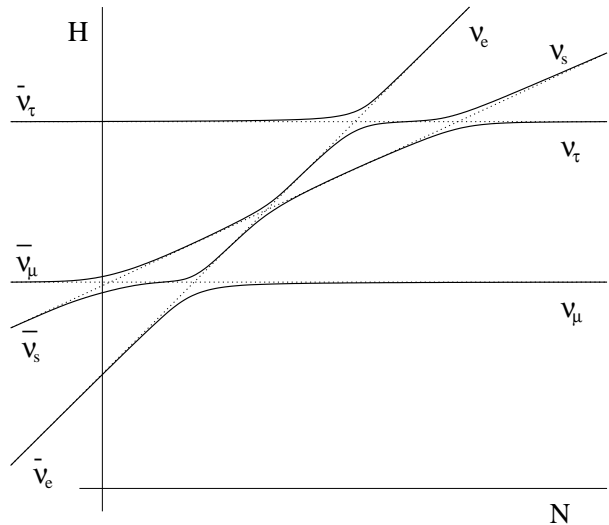}\hfil}
\caption{~~The level crossing pattern for the intermediate mass 
scale scenario. 
}
\end{figure}
%%%%%%%%%%%%%%%%%%%%%%%%%%%%%%%%%%%%%%%%%%%%%%%%%%%%%%%%%%%%%%%%%%%
If all the resonances are effective, the following transitions  occur 
inside the star: 
\be
\nu_e \rightarrow \nu_{\tau},~~~
\nu_{\mu} \rightarrow \nu_e,~~~
\nu_{\tau} \rightarrow \nu_4' \rightarrow  
\nu_e  \rightarrow \nu_2  (\nu_{\mu}, \nu_s)~. 
\ee
Thus at the Earth  one expects: 
(i) disappearance of the
$\nu_e$ neutronization peak
(appearance of the peak in non-electron neutrinos); 
(ii) the electron neutrinos with the  hard spectrum
(of original $\nu_{\mu}$); 
(iii) tau neutrinos with  soft spectrum (corresponding 
initial $\nu_e$).\\

If resonances $\nu_e - \nu_2$ and $\nu_e - \nu_4$ (at low densities) 
are inefficient, the transitions proceed as: 
\be
\nu_e \rightarrow \nu_{\tau},~~~
\nu_{\mu} \rightarrow \nu_2~~ (\nu_{\mu}, \nu_s),~~~
\nu_{\tau} \rightarrow \nu_e  \rightarrow \nu_4 ~~ (\nu_{\mu}, \nu_s). 
\ee
Thus, the  
neutronization peak changes the flavor, 
the flux of $\nu_e$ in the cooling stage is strongly suppressed. 
Non-electron neutrinos will have soft component.  

If $\nu_e - \nu_{\tau}$ resonance is inefficient 
(because of smallness of $U_{e3}$), the following transitions 
occur: 
\be
\nu_e \rightarrow \nu_4 (\nu_{\mu}, \nu_s),~~~
\nu_{\mu} \rightarrow \nu_e,~~~
\nu_{\tau} \rightarrow \nu_2  (\nu_{\mu}, \nu_s).
\ee
In this case the $\nu_e$-flux is unchanged in whole region 
of $r$-processes. At the detector, however, 
$\nu_e$ will have a hard spectrum. The non-electron neutrino 
spectrum will have both hard and soft components. 

%%%%%%%%%%%%%%%%%%%%%%%%%%%%%%%%%%%%%%%%%%%%%%%%%%%%%%%%%%%%%%%%%%%%%%%
%%%%%%%%%%%%%%%%%%%%%%%% ss5.6 %%%%%%%%%%%%%%%%%%%%%%%%%%%%%%%%%%%%%%%%
\subsection{Scheme with two degenerate neutrinos}
%%%%%%%%%%%%%%%%%%%%%%%%%%%%%%%%%%%%%%%%%%%%%%%%%%%%%%%%%%%%%%%%%%%%%%%

Maximal  mixing prefers  strong mass  degeneracy. Therefore 
the  atmospheric neutrino result \cite{SK-atm} 
can be considered as an indication that 
$\nu_{\mu}$ and $\nu_{\tau}$  are strongly mixed in the  
two heavy almost  degenerate neutrino states: 
$\Delta m \ll   m_2 \approx m_3 \approx m_0$. 
If $m_0 \sim 1$ eV,  these neutrinos can compose the 
2$\nu$ HDM component in the Universe. In this case 
$\Delta m \approx (2 - 5)\times 10^{-3}$ eV. The first  neutrino 
composed, mainly,  of   $\nu_e$  
can be much lighter: $m_1 \ll m_0$,  so that no observable signal in  
the double beta decay is expected. 
To explain the solar neutrino deficit one can introduce 
sterile neutrino which mixes with $\nu_e$. 
Then solar neutrinos can undergo the $\nu_e \rightarrow \nu_s$ 
resonance conversion.   
The scheme (fig.~12) can also explain the LSND 
result if the admixture of the $\nu_e$ in the heavy state 
is large enough $U_{e3} \sim 2\times 10^{-2}$ \cite{four}.  
%%%%%%%%%%%%%%%%%%%%%%%%%%%%%%%%%%%%%%%%%%%%%%%%%%%%%%%%%%%%%%%%%%%
%%%%%%%%%%%%%%%%%%%%%%%%%%%%% deg.eps %%%%%%%%%%%%%%%%%%%%%%%%%%%%%
\begin{figure}[htb]
\hbox to \hsize{\hfil\epsfxsize=9cm\epsfbox{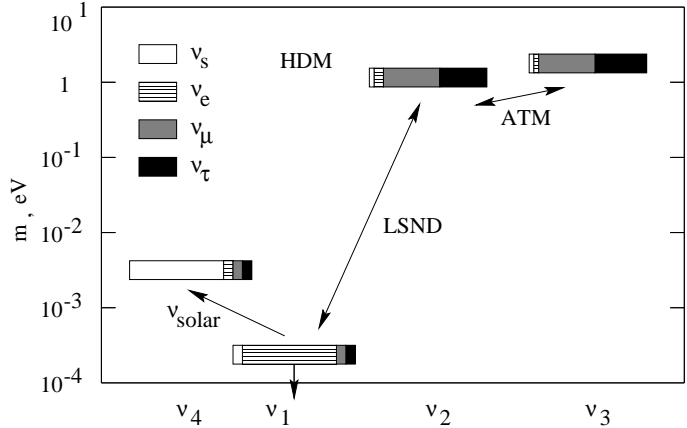}\hfil}
\caption{~~The pattern of the neutrino mass and mixing in the 
scheme with two degenerate neutrinos and one sterile 
component. 
}  
\end{figure}
%%%%%%%%%%%%%%%%%%%%%%%%%%%%%%%%%%%%%%%%%%%%%%%%%%%%%%%%%%%%%%%%%%%
 
With mixing required by the solar neutrino data and the LSND 
result both $\nu_e - \nu_s$ and $\nu_e - \nu_{\tau}$ 
resonances are in the adiabatic domain for  
supernovae.  If all level crossings  are adiabatic,  
then according to the level crossing scheme of fig.~12  
one expects transitions: 
\be   
\nu_e \rightarrow \nu_3' (\nu_{\mu}, \nu_{\tau}),~~~
\nu_2' (\nu_{\mu}, \nu_{\tau}) \rightarrow  \nu_s  \rightarrow
\nu_e ,~~~
\nu_3' (\nu_{\mu}, \nu_{\tau}) \rightarrow \nu_e  \rightarrow \nu_s.
\label{degnu}
\ee
As the consequence, 
(i) the neutronization peak  changes flavor; 
(ii) the electron neutrinos at the cooling stage 
have hard spectrum due to 
spectra interchange; (iii) $\nu_s$ flux  appears, and 
therefore total flux of  the active neutrinos  decreases;    
(iv) no modification of the $\bar{\nu}_e$-spectrum is expected. 

Due to $\nu_{ne} \rightarrow \nu_e$ conversion in the high density
resonance the $\nu_e$- flux with hard spectrum appears 
in the outer part of the r-processes region. 
This will prevent from desired nucleosynthesis. 
The problem  can be avoided if  $\nu_s$ admixture in the heaviest state 
is absent. 
Then, 
one of combinations $(\nu_{\mu}, \nu_{\tau})$, $\nu_3'$ will transfer 
in to another combination of the same components $\nu_2'$, 
in turn $\nu_2'$ is  transformed into sterile neutrino, 
so that the $\nu_e$-flux in the outer part of the $r$-process
region will be absent. 
%%%%%%%%%%%%%%%%%%%%%%%%%%%%%%%%%%%%%%%%%%%%%%%%%%%%%%%%%%%%%%%%%%%%
%%%%%%%%%%%%%%%%%%%%%%%%%%%%  degenerate %%%%%%%%%%%%%%%%%%%%%%%%%%%
\begin{figure}[htb]
\hbox to \hsize{\hfil\epsfxsize=9cm\epsfbox{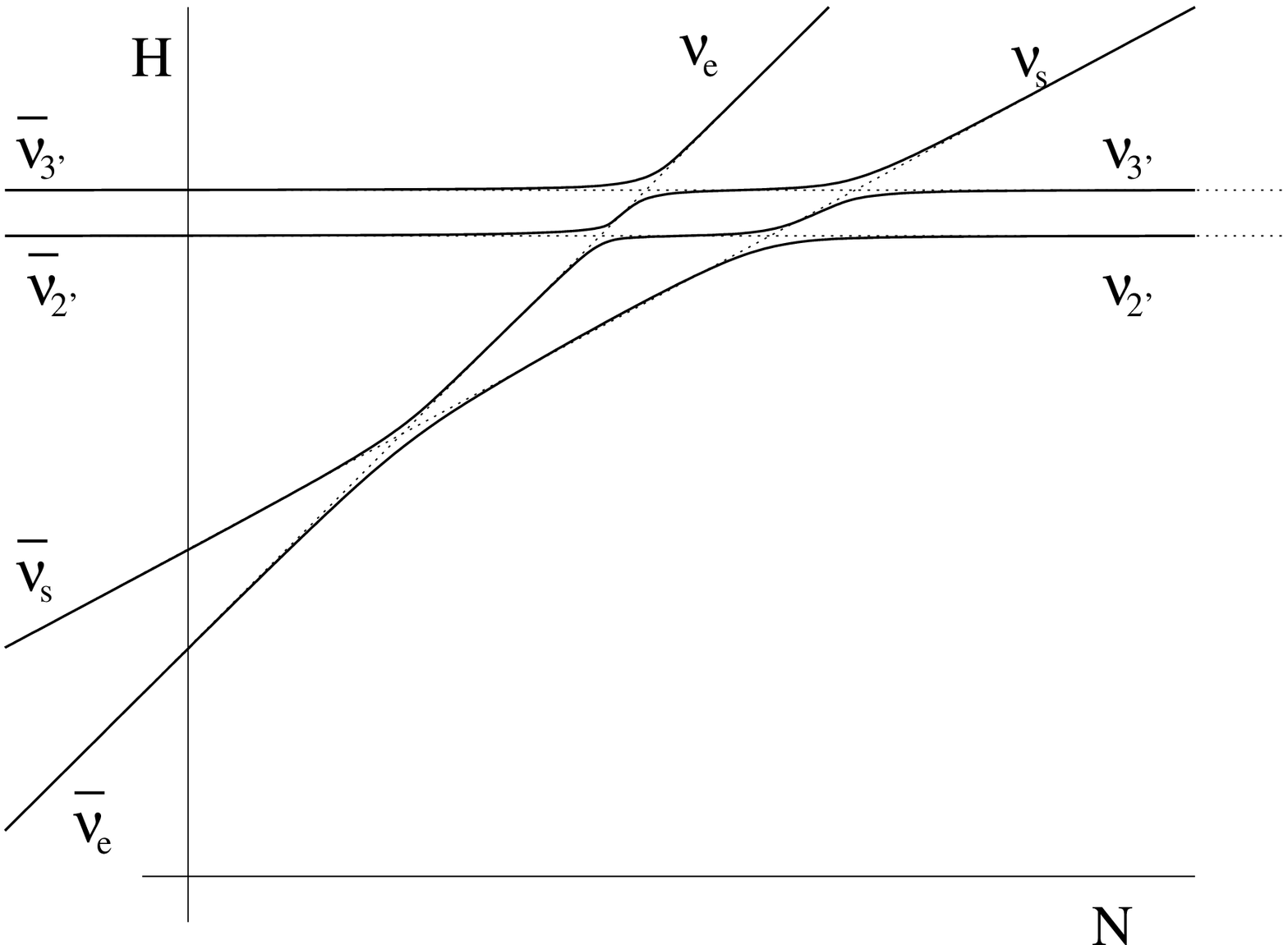}\hfil}
\caption{~~The level crossing pattern in the scheme 
with two degenerate neutrinos 
and one sterile component. 
}
\end{figure}
%%%%%%%%%%%%%%%%%%%%%%%%%%%%%%%%%%%%%%%%%%%%%%%%%%%%%%%%%%%%%%%%%%%%
The fluxes at the Earth will be similar to those in the 
previous case (\ref{degnu}). The difference is that now there is no
$\nu_s$ flux at the exit, and the total flux of the  
active neutrinos is unchanged.  

%%%%%%%%%%%%%%%%%%%%%%%%%%%%%%%%%%%%%%%%%%%%%%%%%%%%%%%%%%%%%%%%%%
%%%%%%%%%%%%%%%%%%%%%%%%%%%%%% ss5.7  %%%%%%%%%%%%%%%%%%%%%%%%%%%%
\subsection{Grand Unification Scenario}
%%%%%%%%%%%%%%%%%%%%%%%%%%%%%%%%%%%%%%%%%%%%%%%%%%%%%%%%%%%%%%%%%%%%

The see-saw mechanism  based on the
Grand Unification   leads to the mass of the heaviest
neutrino ($\approx \nu_{\tau}$) in
the range $(2 - 3)\cdot 10^{-3}$ eV, and hence,  to a solution of the
solar
neutrino problem through the $\nu_e \rightarrow \nu_{\tau}$ conversion.
An  existence of the  light singlet fermion, $\nu_s$,  which
mixes predominantly with muon neutrino through the mixing mass
$m_{\mu s} \sim O(1)$ eV 
allows one \cite{JS} (i) to solve the atmospheric neutrino problem
via the $\nu_{\mu} \leftrightarrow \nu_s$ oscillations, (ii) to
explain the LSND result and (iii) to get two component
hot dark matter in the Universe (fig.~14). Similar scheme 
has been suggested previously in another context \cite{GU}. 
%%%%%%%%%%%%%%%%%%%%%%%%%%%%%%%%%%%%%%%%%%%%%%%%%%%%%%%%%%%%%%%%%%%%
%%%%%%%%%%%%%%%%%%%%%%%%%%%%%%%%%%%%%%%%%%%%%%%%%%%%%%%%%%%%%%%%%%%%
\begin{figure}[htb]
\hbox to \hsize{\hfil\epsfxsize=9cm\epsfbox{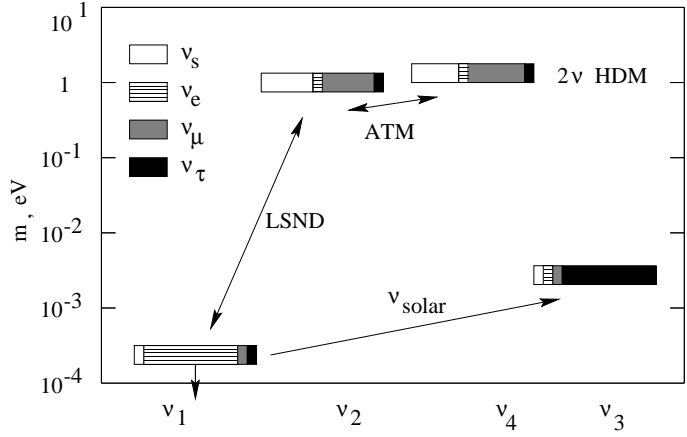}\hfil}
\caption{~~The pattern of the neutrino mass and mixing 
in the Grand Unification scenario. 
}
\end{figure}
%%%%%%%%%%%%%%%%%%%%%%%%%%%%%%%%%%%%%%%%%%%%%%%%%%%%%%%%%%%%%%%%%%%%%%%

The level crossing pattern  can be found in the following 
way (fig.~15).
One diagonalizes first the strongly mixed heavy sub-system
$\nu_{\mu}$ and $\nu_s$. This gives the levels
$\nu_{2m}'$, $\nu_{4m}'$. Then using smallness of all other mixings
one gets the level crossings (resonances):
$\nu_e - \nu_{4m}'$,  $\nu_e - \nu_{2m}'$
at large densities,  and $\nu_e -  \nu_{\tau}$ crossing at small
density. There is no level crossing in the antineutrino channels.
%%%%%%%%%%%%%%%%%%%%%%%%%%%%%%%%%%%%%%%%%%%%%%%%%%%%%%%%%%%%%%%%%%
%%%%%%%%%%%%%%%%%%%%%%%%%%%%% GUT.eps %%%%%%%%%%%%%%%%%%%%%%%%%%%%
\begin{figure}[htb]
\hbox to \hsize{\hfil\epsfxsize=8cm\epsfbox{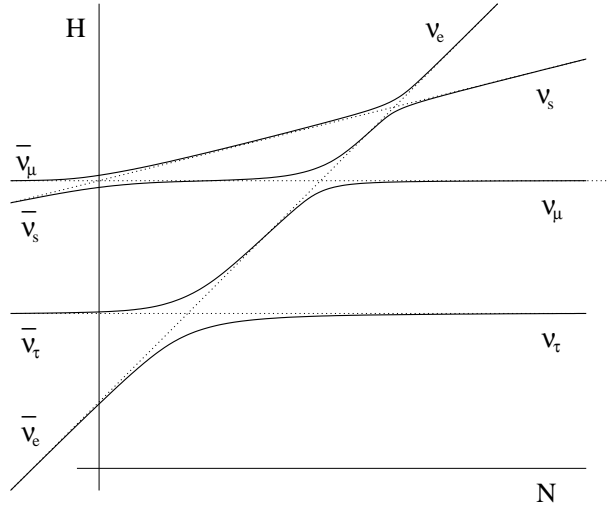}\hfil}
\caption{~~The level crossing pattern in the Grand Unification scenario. 
}
\end{figure}
%%%%%%%%%%%%%%%%%%%%%%%%%%%%%%%%%%%%%%%%%%%%%%%%%%%%%%%%%%%%%%%%%%%%%

If the adiabaticity condition is fulfilled
in all the resonances, one predicts
the following transitions:
\begin{equation}
\nu_e \rightarrow \nu_{4m}' \approx (\nu_{\mu} + \nu_s)/\sqrt{2},~~~~ 
\nu_{\mu} \rightarrow \nu_e \rightarrow \nu_{\tau},~~~~ 
\nu_{\tau}  \rightarrow \nu_e.
\end{equation}
Thus,  one can observe the electron neutrinos with  hard spectrum, 
$\nu_{\mu}$'s with the soft spectrum
and the flux of sterile neutrinos being 
about 1/12 of the total  neutrino flux. 
Half of the $\bar{\nu}_{\mu}$-flux will be converted to 
the $\nu_s$-flux. 
The production of the heavy elements  due to the $r$ - processes in 
supernovae imply that transitions
$\nu_{\tau} \rightarrow \nu_e$,  $\nu_{\mu} \rightarrow \nu_e$
are suppressed in the inner parts of the star
\cite{r-proc}. As follows from the 
level crossing scheme  the appearance of $\nu_e$ can be
due to adiabatic transition  $\nu_{\mu} \rightarrow \nu_e$.
The problem can be solved if 
$\Delta m^2 \approx m_{\mu s}^2 < (1 -  2)$ eV$^2$, so that  
the transitions occur in the outer layers above the 
$r$-processes region. 

%%%%%%%%%%%%%%%%%%%%%%%%%%%%%%%%%%%%%%%%%%%%%%%%%%%%%%%%%%%%%%%%%%%%%%
%%%%%%%%%%%%%%%%%%%%%%%%%% ss5.8  %%%%%%%%%%%%%%%%%%%%%%%%%%%%%%%%%%%%%
\subsection{Matter induced resonance conversion}
%%%%%%%%%%%%%%%%%%%%%%%%%%%%%%%%%%%%%%%%%%%%%%%%%%%%%%%%%%%%%%%%%%%%%%%
Previous analysis of transitions was based on assumption that 
no sterile neutrino flux is produced in the central regions of a star. 
This may not be true if $\nu_e - \nu_s$ mixing mass term is large enough. 
Indeed, the effective potential for $\nu_e - \nu_s$ channel 
(\ref{ester}) equals zero at 
\be
n_n = 2 n_e~. 
\ee 
This condition is satisfied in  the layer  
with significant neutronization with $\rho \sim 10^{11}$ 
g/cm$^3$. Since $\Delta m^2/2E \ll \sqrt{2} G_F \rho/m_N $ 
in this layer, the resonance condition takes
the form $V_{es} \approx 0$ \cite{voloshin}. Moreover, this condition is 
satisfied both for neutrinos and antineutrinos.   
Thus,  at  $V_{es} \approx 0$ the resonance conversions   
${\nu}_e \rightarrow {\nu}_s$ and  
$\bar{\nu}_e \rightarrow \bar{\nu}_s$ occur.

Thus,  the level crossing schemes of 
Figs. 11, 13, 15   should be completed 
by two more resonances  at 
densities $\rho \sim 10^{11}$ g/cm$^3$  in the neutrino and
in antineutrino channels. 
%%%%%%%%%%%%%%%%%%%%%%%%%%%%%%%%%%%%%%%%%%%%%%%%%%%%%%%%%%%%%%%%%
%%%%%%%%%%%%%%%%%%%%%%%%%%%%% matter.eps %%%%%%%%%%%%%%%%%%%%%%%%%%%%
\begin{figure}[htb]
\hbox to \hsize{\hfil\epsfxsize=8cm\epsfbox{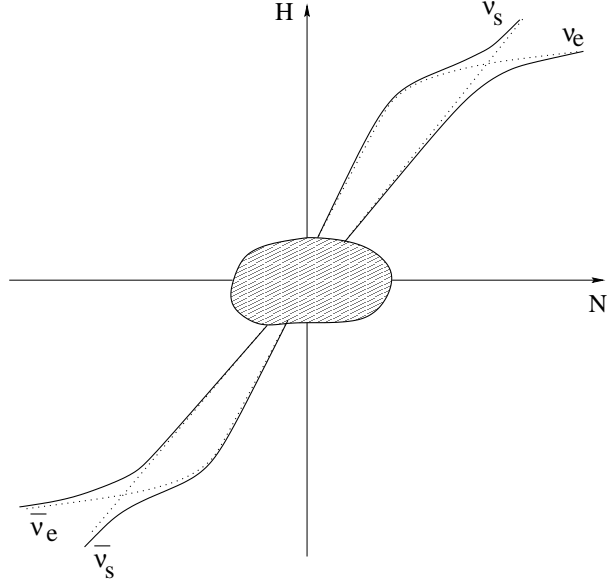}\hfil}
\caption{~~The $\nu_e - \nu_s$ and
$\bar{\nu}_e - \bar{\nu}_s$ level crossings 
in central parts of supernova. 
Dashed lines show the dependences of $\nu_e$ and  $\nu_s$ energies 
on the total density.  
The shadowed region corresponds to the level crossing patterns 
shown in figs. 11, 13, 15.  
}
\end{figure}
%%%%%%%%%%%%%%%%%%%%%%%%%%%%%%%%%%%%%%%%%%%%%%%%%%%%%%%%%%%%%%%%%%%%%

Efficiency of transition in resonances is determined by 
the adiabaticity condition which in turn depends on mixing. 
The condition can be rewritten as: 
\be
\sin^2 2\theta > \frac{2E}{\Delta m^2 r_V}~,  
\ee 
where $r_V \equiv V/dV/dr$ is the scale height of the change of the 
effective potential. For $r_V \sim 10$ km and $E \sim 10$ MeV we get 
\be
\sin^2 2\theta > 10^{-2} \left(\frac{ 1 {\rm eV}}{m}\right)^2~.  
\label{sadiab}
\ee
If $\nu_s$ mixes only in the light states ({\it e.g.} to explain the 
solar neutrino deficit),  then $m \sim 3 \times 10^{-3}$ eV and  clearly 
the adiabaticity condition is not satisfied.  
The situation  can be different, when  there is some admixture 
of $\nu_s$ in the heavy mass eigenstates with $m >  1$ eV. 
For instance,  in the Grand Unification scenario  (fig.~14)  the 
$\nu_e$ admixture in $\nu_2$ or $\nu_4$ 
required by  the LSND 
data is enough to satisfy the adiabaticity condition (\ref{sadiab}) and
therefore 
to induce strong $\nu_e \rightarrow \nu_s$ and 
$\bar{\nu}_e \rightarrow \bar{\nu}_s$ transitions in central parts of the
star. Both transitions have practically the same efficiency.

Notice that $\bar{\nu}_e \rightarrow \bar{\nu}_s$ transition leads 
to disappearance of the $\bar{\nu}_e$ signal which is crucial for 
the present searches of the $\nu$-bursts from supernovae. 
The observation of $\bar{\nu}_e$  signal from SN87A 
gives the bound on $\bar{\nu}_e \rightarrow \bar{\nu}_s$ transition and 
therefore on mixing of $\nu_e$ in the heavy state. 
At the same time, an efficiency of the  transition depends  
on model of the star, and in particular,  on its  mass. 
For some class of  stars, like SN87A, the transition can be 
less efficient leading to partial (weak) suppression of signal. 
For other cases the transition can be strong. This is 
clearly, important for $\nu$-burst  detection. 

Notice also that transitions  
$\nu_e \rightarrow \nu_s$ and 
$\bar{\nu}_e \rightarrow \bar{\nu}_s$ 
lead  to disappearance of the $\nu_e$-flux 
in the inner  part  of the $r$-processes region  
and disappearance of the $\bar{\nu}_e$-flux  in  whole the region.

%%%%%%%%%%%%%%%%%%%%%%%%%%%%%%%%%%%%%%%%%%%%%%%%%%%%%%%%%%%%%%%%%%%%%%%%%
%%%%%%%%%%%%%%%%%%%%%%%%%%%%%%%%%%%%%%%%%%%%%%%%%%%%%%%%%%%%%%%%%%%%%%%%%

\section{Conclusions}

Variety of physical conditions,  the effective 
density profiles and still  possible neutrino mass 
spectra 
leads to a variety of possible neutrino conversion phenomena.

The picture of the neutrino transformations  
depends significantly on scheme of neutrino mass and mixing. 
The medium effects are minimal in the case of strict bi-maximal 
mixing. In this case,  matter can only  suppress mixing in the 
$\nu_e$-channels of oscillations, and still  the results are the
same as in the case of averaged vacuum oscillations. In contrast, 
there is a richness of the matter effects in   
schemes with small mixing and especially in schemes with 
sterile neutrinos. 

The study of supernova neutrinos will allow one  
to test whole  spectrum of neutrino masses.

\bigskip

{\bf Acknowledgement}\\

I would like  to thank H. Minakata and O. Yasuda 
for hospitality during my stay at TMU. 
The author is  grateful to A. Dighe,  H. Minakata and F. Vissani for
fruitful discussions and comments.

\end{document}